\documentclass[12pt]{article}
\usepackage{amsmath}
\usepackage{graphicx}
\usepackage{enumerate}
\usepackage[round]{natbib}
\usepackage{url} 
\usepackage[top=1in, bottom=1in, left=1in, right=1in]{geometry}
\usepackage{color, amsmath, amssymb, amsbsy, amsthm, graphicx, bbm, amsfonts, bm,dsfont,enumitem}
\usepackage{setspace, comment, graphicx, algorithm, latexsym,titlesec}
\usepackage[flushleft]{threeparttable}
\usepackage[colorlinks,allcolors=blue]{hyperref} 
\usepackage{titlesec}
\allowdisplaybreaks
\usepackage{algorithm,algpseudocode}
\usepackage{wrapfig,lipsum,booktabs}
\usepackage{multirow}

\newtheorem{theorem}{Theorem}

\newtheorem{condition}{Condition}
\newtheorem{assumption}{Assumption}

\newcommand{\Expectation}{\mathbb{E}}

\newcommand{\Prob}{\mathbb{P}}
\newcommand{\Indicator}{\mathds{1}}
\newcommand{\op}{o_{\mathrm{p}}}

\newcommand\Independent{\protect\mathpalette{\protect\independenT}{\perp}}
\newcommand{\Independence}{\Independent}
\def\independenT#1#2{\mathrel{\rlap{$#1#2$}\mkern2mu{#1#2}}}
\newcommand{\toProb}{\overset{\mathrm{p}}{\to}}
\newcommand{\toDist}{\overset{\mathrm{d}}{\to}}

\DeclareMathOperator*{\argmin}{argmin}

\newcommand\blfootnote[1]{%
  \begingroup
  \renewcommand\thefootnote{}\footnote{#1}%
  \addtocounter{footnote}{-1}%
  \endgroup
}

\begin{document}

\begin{center}
\Large Covariate-Adjusted Response-Adaptive Design with Delayed Outcomes\blfootnote{\hskip -1.7em The authors are in alphabetical order.
}
\end{center}

\vskip 2em

\begin{center}
\begin{tabular}{ccc}
Xinwei Ma & Jingshen Wang & Waverly Wei \\ 
&& \\
\footnotesize Department of Economics & \footnotesize Division of Biostatistics & \footnotesize Department of Data Sciences  \\
\footnotesize  & \footnotesize  & \footnotesize  and Operations \\
\footnotesize University of California San Diego & \footnotesize University of California Berkeley & \footnotesize University of Southern California
\end{tabular}
\end{center}

\vskip 1em

\begin{center}
\footnotesize \today \\
\end{center}

\vskip 3em
\onehalfspacing
\begin{abstract}
\noindent Covariate-adjusted response-adaptive (CARA) designs have gained widespread adoption for their clear benefits in enhancing experimental efficiency and participant welfare. These designs dynamically adjust treatment allocations during interim analyses based on participant responses and covariates collected during the experiment. However, delayed responses can significantly compromise the effectiveness of CARA designs, as they hinder timely adjustments to treatment assignments when certain participant outcomes are not immediately observed. In this paper, we propose a fully forward-looking CARA design that dynamically updates treatment assignments throughout the experiment as response delay mechanisms are progressively estimated. Our design strategy is informed by novel semiparametric efficiency calculations that explicitly account for outcome delays in a multi-stage setting. Through both theoretical investigations and simulation studies, we demonstrate that our proposed design offers a robust solution for handling delayed outcomes in CARA designs, yielding significant improvements in both statistical power and participant welfare.

\vskip 1.5em

\noindent\textit{Key\hskip0.05em words}: Delayed outcomes; Frequentist adaptive experimental design; Response adaptive designs.
\end{abstract}

\vfill

\newpage

\onehalfspacing

\section{Introduction}\label{Section-1: Motivation and contribution}

\subsection{Motivation and Contribution}

In clinical trials, social science field experiments, and A/B tests, participants often enroll sequentially, and responses to treatments can vary due to individual-specific characteristics. In these contexts, covariate-adjusted response-adaptive (CARA) designs have gained increasing popularity: by dynamically adjusting treatment assignments based on accumulated covariates and outcome information during the experiment, CARA designs can be adopted to optimize treatment allocation to improve statistical power \citep{hahn2011adaptive, hu2015unified, blackwell2022batch}, reduce participant exposure to less effective treatments and enhance overall welfare \citep{hu2006theory, hu2015unified, wei2024fair, wei2025adaptive}, and maintain statistical validity for analyzing experimental data \citep{zhang2007asymptotic, hu2012asymptotic, bugni2018inference, offer2021adaptive, bai2022inference, robertson2023response, zhao2023adaptive, wei2025adaptive, bibaut2025demystifying}. 

Nevertheless, because CARA designs adjust treatment assignments based on observed participant responses, delays in outcome observation can significantly hinder their effectiveness. Intuitively, if primary outcomes for some participants remain unobserved at the time of interim analysis, the experimental designer may lack information needed to optimize treatment assignments for future participants with similar covariate profiles. These delays often depend on both the treatment arm and participant covariates, as is common in sequentially enrolled randomized experiments. 

As an example, we revisit the study by \cite{fahey2020financial}, a randomized experiment conducted in the Shinyanga region of Tanzania between April 24 and December 14, 2018. The study aimed to assess the impact of cash incentives on retention in care and viral suppression among people living with HIV (PLHIV). A total of 530 PLHIV aged 18 or older were sequentially enrolled and randomly assigned to either a treatment group receiving cash incentives or a control group without any incentive. Viral load was measured six months after treatment assignment through blood draws. However, delays occurred, as the lab test required a separate clinical visit and some participants missed their scheduled appointments. We calculated the number of days delayed as the difference between the date of the viral load test and the six-month follow-up date, separately for two covariate strata defined by biological sex. From Figure \ref{Fig-1: empirical example, delay distribution, X = sex}, it is clear that the distribution of outcome delays varies significantly depending on treatment assignment and gender. Specifically, the treatment arm exhibits less delay, and within the treatment arm, the delayed response issue seems less pronounced among female participants. Overall, the empirical observation highlights the importance of accounting for delayed outcomes that are both arm- and covariate-specific.

\begin{figure}[h]
\centering
\includegraphics[width=0.7\linewidth]{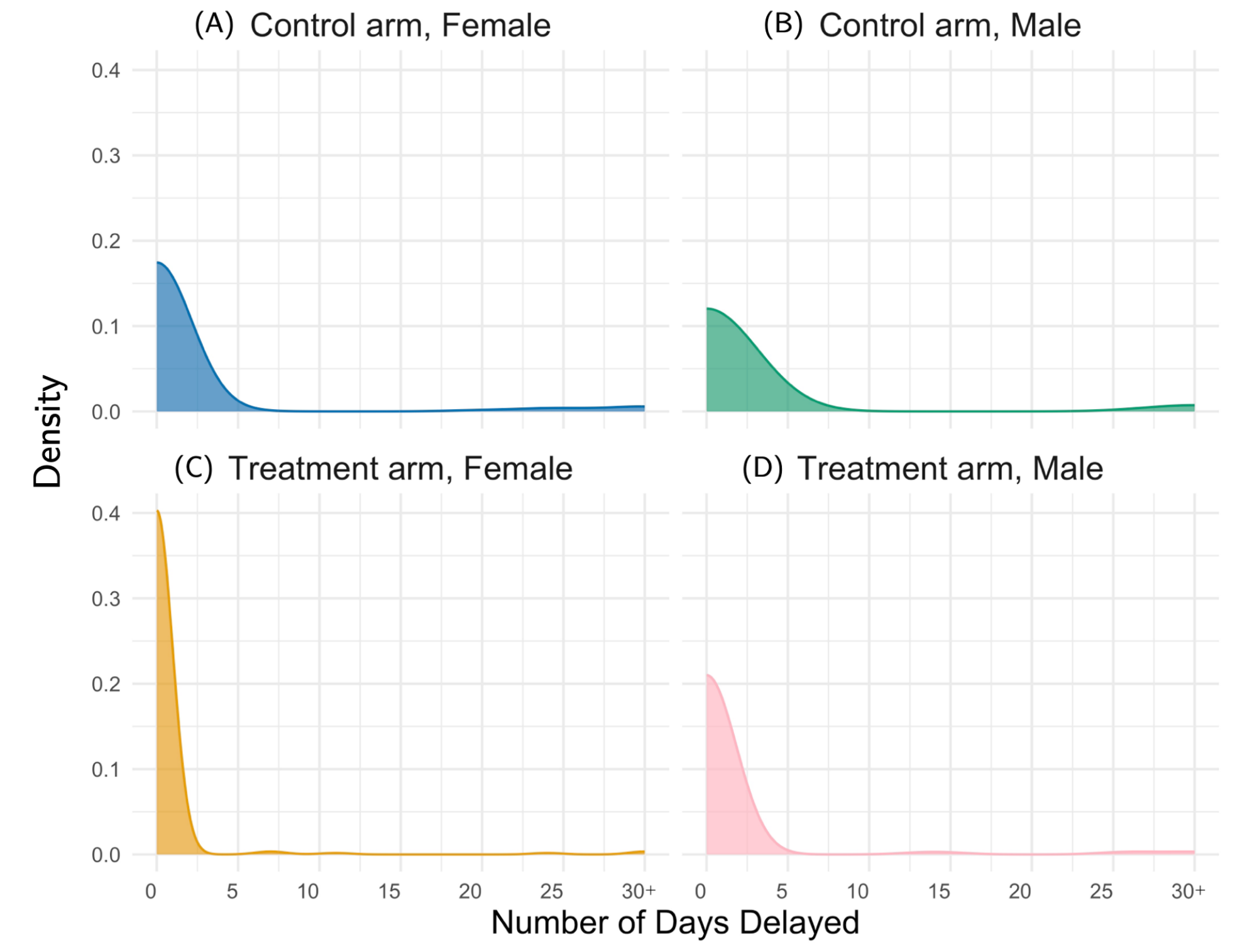}
\caption{Delays in viral load blood test across treatment arms and covariate strata (biological sex). 
}
\label{Fig-1: empirical example, delay distribution, X = sex}
\end{figure}

To our knowledge, although there has been prior work on developing valid statistical inference procedures in the presence of delayed responses in CARA \citep{bai2002asymptotic, hu2008doubly,robertson2023response}, no CARA designs has incorporated the delay mechanism when optimizing experimental objectives. In this paper, we address the challenges that response delays pose to covariate-adjusted response-adaptive (CARA) designs. We begin by examining whether existing optimal allocation strategies remain valid under delayed outcomes and, when they do not, derive the optimal allocation in such settings. We then develop methods to adapt CARA designs accordingly. Finally, we propose an efficient estimator that accounts for delayed responses at the end of the experiment. We elaborate on these contributions in more detail below.

First, we demonstrate that the classical optimal allocation strategy used in existing CARA designs no longer provides optimal treatment assignment when accounting for arm- and covariate-dependent delays. Specifically, we first derive a semiparametric efficiency bound for treatment effect estimation in the presence of outcome delay (Theorem \ref{Thm-1: Semi-parametric efficiency bound}), which differ substantially from existing results in the literature \citep{hahn1998role,hirano2003efficient, cattaneo2010efficient, zhu2023covariate, wu2025nonstationary}. See \cite{van2000asymptotic} and \cite{tsiatis2006semiparametric} for general discussions on semiparametric efficiency. This result allows us to show that the optimal delay mechanism involves a unique trajectory of treatment assignment probabilities that vary across different stages, resulting in treatment allocations with lower variance for power maximization, and enhanced failure reduction in welfare improvement designs. To our knowledge, this is the first design to explicitly explore delay mechanisms to optimize trial objectives, while existing literature focuses on conducting sequential tests \citep{hampson2013group}, designing Bayesian adaptive designs where delay is independent of the arm and covariates \citep{lin2020adaptive}, or conducting statistical inference after experiments have concluded \citep{bai2002asymptotic,hu2008doubly}. 

Second, since the delay mechanism is unknown a priori and remains only partially estimable during the experiment, we propose a fully forward-looking CARA design that sequentially updates treatment assignment probabilities to target a trial objective at the conclusion of the experiment. In this process, our algorithm offers several extrapolation strategies when estimating delay mechanism, allowing better alignment with the designer’s prior knowledge. We further discuss in detail two design objectives frequently adopted in current CARA literature: maximizing experiment power \citep{tymofyeyev2007implementing, zhao2023adaptive} and improving overall participant welfare (or failure reduction) while maintaining statistical power constraints \citep{rosenberger2001optimal, robertson2023response}. We justify the benefits of our proposed forward-looking CARA design through theoretical investigation (Section \ref{Section-4: Theoretical investigation}) and a synthetic case study (Section \ref{Section-5: Synthetic case study}). Under the design objective of maximizing experiment power, our proposed forward-looking CARA design attains higher estimation efficiency compared to the Neyman allocation and complete randomization designs in the presence of delayed responses. Under the design objective of improving overall participant welfare, we demonstrate that our proposed forward-looking CARA design achieves additional failure reduction, leading to greater welfare improvement compared to both the ethical design and the complete randomization design while maintaining statistical power. For both design objectives, our proposed design can provide valid statistical inference, which is justified in Theorem \ref{Thm-3: Statistical inference}.

Third, we make theoretical contributions to the literature on CARA and adaptive experimental designs by establishing general conditions under which estimated treatment effects are consistent and asymptotically normally distributed (Theorem \ref{Thm-2: Asymptotic normality of estimated treatment effects}). Specifically, we do not restrict the potential outcomes or the delay mechanism to any parametric family of distributions, thereby alleviating the burden of choosing a particular set of parametric assumptions. Our theoretical development builds on martingale methods \citep{hall2014martingale} and a high-level condition (Condition \ref{Cond-1: Convergence of the optimized treatment allocation}) that the optimized allocations converge in large samples, an assumption satisfied by many adaptive experimental designs. As a result, the general consistency and asymptotic normality findings may be of independent interest. We then specialize in our proposed design and verify the design consistency condition (Section \ref{section: Convergence of the optimized treatment allocation}), providing a suite of tools for designing and analyzing adaptive experiments with delayed outcomes, along with rigorous statistical guarantees.

\subsection{Setup: CARA with Arm- and Covariate-dependent Delays}\label{Section-1-1:Setup: CARA with arm- and covariate- dependent delays} 

In the remainder of this introduction, we lay out the setup of CARA when responses are subject to delay. Formal statements of our assumptions are collected in Section \ref{Section-4: Theoretical investigation}. The experiment is conducted over $T$ stages, labeled  $t = 1, 2, \dots, T$. In each stage $t$, $n_t$ participants are enrolled. Participants are indexed by $i = 1, 2, \dots, N$,  with $N = \sum_{t=1}^T n_t$ being the total sample size. We use $N_t$ to denote the cumulative sample size at the end of stage $t$, that is, $N_t = \sum_{s=1}^t n_s$. Therefore, $N= N_T$, where we omit the subscript to save notation. Finally, $S_i = t$ denotes that individual $i$ is enrolled in stage $t$. 

Upon enrollment, baseline covariates $X_{i} \in \mathcal{X}$ are collected for each participant, and they are then randomized into one of the treatment arms, denoted by $ A_{i} \in \mathcal{A} = \{0,1\}$, which represents their actual treatment status. Following the Neyman-Rubin causal model, the potential outcomes are denoted by $Y_{i}(0)$ and $Y_i(1)$. In line with existing literature in adaptive design, we are in a scenario where the potential outcomes and the covariates, $(Y_{i}(0), Y_{i}(1), X_{i})$, are independent and identically distributed across different stages. For future reference, we also introduce notation for mean potential outcomes: $\mu(x,a) = \Expectation[Y_i(a)|X_i=x]$ and $\mu(a) = \Expectation[Y_i(a)]$, and the conditional and unconditional treatment effects are defined as
$\tau(x) = \mu(x,1) - \mu(x,0)$ and $ \tau = \mu(1) - \mu(0)$. Other moments of the potential outcomes are $\mu_s(x,a) = \Expectation[Y_i(a)^s|X_i=x]$ and $\mu_s(a) = \Expectation[Y_i(a)^s]$, but we omit the subscript whenever $s=1$ for ease of notation.

Since outcome information may not be immediately available after treatment assignment, we use $D_{i}$ to represent the number of stages after which the experimenter can observe $Y_{i}$. This means that the participant outcome $Y_i$ is observed at the end of stage $t$ if and only if $D_i + S_i \leq t$. To give a concrete example, Let $S_i=1$ so the participant is enrolled in the first stage. Then, their outcome information is available at the end of stage 3 if and only if the delay is no more than two stages, which means $D_i\leq 2$, or equivalently, $S_i+D_i \leq 3$. Also note that when $D_i=0$, it implies no delay in the outcome for this individual. 

The delay distribution is denoted by $\rho(d|x,a) = \Prob[D_i \leq d|X_i=x,A_i=a]$. As the notation suggests, we allow the delay mechanism $D_{i}$ to depend on the covariates and the treatment assignment status. We will assume delay is conditionally independent of the potential outcomes: $D_{i} \Independent (Y_i(0), Y_i(1), S_i)\  \mid\  (X_{i}, A_{i})$. As the delay distribution does not shift over time, it allows us to sequentially learn the delay mechanism using accrued data and to optimize treatment allocation. 

In CARA designs, the treatment assignment is sequentially updated based on accumulated data to achieve a pre-specified design objective. Specifically, at stage \( t \), suppose we have collected historical participant information, denoted as $ \mathcal{H}_{t} = \{ (A_{i}, X_{i}, Y_{i}, D_{i}) \}_{i=1}^{N_t}$, 
where we recall that the sample indexed by $i=1,2,\dots,N_t$ consists of all participants enrolled in the first $t$ stages. The treatment assignment for the next stage,  \( t+1 \), denoted as \( \hat{e}^*_{t+1}(a|x) \), is determined by both this historical information and the covariate information of the newly enrolled participants in Stage $t+1$. Formally, this is captured by $\Prob[A_{i} = a|X_{i}=x,S_i=t+1,\mathcal{H}_{t}] = \hat{e}^*_{t+1}(a|x)$, which means the treatment assignment adapts to $\mathcal{H}_{t}$.

In this manuscript, we work under the setting that $T$ and $n_t$ are pre-fixed. We also hope to note that, in general adaptive trial designs, determining the appropriate sample size and the number of stages involves balancing statistical rigor, operational feasibility, and ethical considerations. Before an experiment starts, practitioners often pre-specify the desired power and significance level, along with the expected treatment effect size based on prior studies or clinical knowledge, to compute an initial overall sample size.  Then, the number of stages is selected by evaluating logistical considerations (e.g., recruitment rate, data availability, cost constraints). It is often the case that fewer stages are used to simplify logistics and reduce operational complexity, while more stages increase flexibility but impose higher management costs. Once the number of stages is decided, sample size allocation per stage can be determined either equally or unequally (e.g., adaptive design with small initial pilot study) based on anticipated interim analyses. It is also common practice to conduct simulations during the planning stage to evaluate various stage/sample size configurations, ensuring feasibility and ethical appropriateness of the chosen adaptive strategy. Our design thus aligns with the case where there is a finite number of stages and each stage contains an equal/unequal number of participants.

\section{Optimal Delay-adjusted Treatment Allocation in CARA}\label{Section-2: Optimal delay-adjusted treatment allocation in CARA}

While the design goals of CARA vary across applications, maintaining sufficient statistical power to assess the treatment's effectiveness on the primary outcome remains a central concern. A statistically efficient estimator of the treatment effect also supports a robust adaptive design, as the estimated effects guide the sequential revision of the treatment allocation strategy. However, it remains unclear what constitutes a good estimator in the presence of delayed outcomes. Moreover, the semiparametric efficiency bound for estimating the treatment effect in multi-stage experiments with delayed responses has not yet been established. To provide practical guidance for designing response-adaptive experiments, we begin by presenting a new result on the semiparametric efficiency bound for estimating the average treatment effect:
\begin{align}\label{Eq-1: semiparametric efficiency bound}
\int\limits_{x\in\mathcal{X}} p(x) \Bigg[\frac{\sigma^2(x,1)}{\sum\limits_{t=1}^T r_t  \rho(T-t|x,1)e_t(1|x)} &+ \frac{\sigma^2(x,0)}{\sum\limits_{t=1}^T r_t  \rho(T-t|x,0)e_t(0|x)} + \big(\tau(x) - \tau\big)^2\Bigg] \mbox{d} x,
\end{align}
where $r_t = \mathbb{P}[S_i=t]$ being the fraction of participants enrolled in stage $t$,  $p(x)$ is the probability density function of the covariate $X$, and $\sigma^2(x,a)$ is the conditional variance of the potential outcome $Y_i(a)$. See Theorem \ref{Thm-1: Semi-parametric efficiency bound} in Section \ref{Section-4: Theoretical investigation} below for a rigorous statement. 

To intuitively understand Eq \eqref{Eq-1: semiparametric efficiency bound}, it helps to first revisit the classical semiparametric efficiency bound in a static (one-period) setting without delays, which is given by:
\begin{align}\label{eq:ATE-sepb}
\int\limits_{x\in\mathcal{X}} p(x) \Bigg[\frac{\sigma^2(x,1)}{ e(1|x) } &+ \frac{\sigma^2(x,0)}{e(0|x) } + \big(\tau(x) - \tau\big)^2\Bigg] \mbox{d} x.
\end{align}
In this classical formulation, the denominators (i.e, propensity scores $e_t(a|x)$) adjust for covariate-specific treatment assignment. 
The key difference between Eq \eqref{eq:ATE-sepb} and Eq \eqref{Eq-1: semiparametric efficiency bound} is the introduction of an additional term $\rho(T-t|x,a)$ in the denominators, which accounts for the probability distribution of response delays. Intuitively, delays introduce extra uncertainty or missingness in observing outcomes. As we operate in a multiple-stage experimental setting, we aggregate these delay-related probabilities across all stages, appropriately weighted by each stage's sample size captured by $r_t$. Hence, the revised efficiency bound appropriately adjusts for both the random assignment of treatments (captured by propensity scores) and the randomness arising from delays in observing responses. 

From the above result, it is evident that when the delay mechanism of the response depends on the covariates and treatment status, the optimal treatment allocation rules in classical CARA designs must be modified to account for outcome delays. In what follows, we examine how the optimal treatment allocation rule is affected by the arm- and/or covariate-dependent delayed responses in two common experimental objectives: one aimed at enhancing statistical power \citep{tymofyeyev2007implementing}, and the other focused on minimizing the expected number of failures while maintaining a minimum power requirement for binary outcomes \citep{rosenberger2001optimal}. In both design goals, the variance of the estimated treatment effects plays an essential role. To simplify the presentation, we focus on two-arm CARA designs with $a\in\{0,1\}$ in this paper. For multi-arm experiments, the design objective can be generalized using non-centrality parameters, as discussed in \cite{lachin1977sample} and \cite{tymofyeyev2007implementing}.\bigskip

\noindent \textbf{Design objective 1: Optimal treatment allocation for power maximization.} The objective is to sequentially assign treatments to minimize the variance of the estimated average treatment effect (ATE), thereby maximizing statistical power. The optimal allocation is the solution to the optimization problem:
\begin{align}\label{Eq-2: Design objective 1 power maximization}
    \min_{e_{1}(\cdot),\dots,e_{T}(\cdot)} \ \mathsf{V}(e_{1}(\cdot),\dots,e_{T}(\cdot)) \quad \text{subject to} \quad \delta \leq e_t(\cdot) \leq 1-\delta \quad \text{for } t=1,\ldots,T,
\end{align}
where $\delta$ ensures that the assignment probabilities are bounded away from 0 and 1, maintaining sufficient randomness in treatment assignments \citep{ma2020robust, heiler2021valid, sasaki2022estimation, ma2022testing, dorn2025much}. The objective function $\mathsf{V}(e_{1}(\cdot),\dots,e_{T}(\cdot))$ is a component of the semiparametric efficiency bound related to the propensity scores, and is defined as:
\[
\mathsf{V}(e_{1}(\cdot),\dots,e_{T}(\cdot)) = \sum_{x \in \mathcal{X}} p(x) \left[ \frac{\sigma_t^2(x,1)}{\sum_{t=1}^T r_t \rho(T-t|x,1)e_t(x) } + \frac{\sigma_t^2(x,0)}{\sum_{t=1}^T r_t  \rho(T-t|x,0)(1-e_t(x))} \right].
\]
In line with existing literature on CARA designs, we explicitly consider discrete covariates in the objective function, as most design applications will first group participants based on their covariate information. Therefore, the objective function takes a summation form with respect to the probability mass function. \bigskip

\noindent \textbf{Design objective 2: Optimal treatment allocation for failure reduction subject to a power constraint.} This design aims to minimize the expected number of failures while satisfying a minimum power requirement for binary responses \citep{rosenberger2001optimal}.  In this context, a treatment is considered to fail when the potential outcome equals one. The optimal allocation thus seeks to assign treatments based on the solution to the following optimization problem:
\begin{align}\label{Eq-3: Design objective 2 failure reduction}
    &\min_{e_{1}(\cdot),\dots,e_{T}(\cdot)}  \ \sum_{t=1}^T r_t \sum_{x\in\mathcal{X}}p(x)\Big[ e_t(x)\left(1 - \mu(x, 1)\right) + \left(1 -e_t(x)\right) \left(1 - \mu(x, 0)\right) \Big]  \quad \\
    &\qquad\qquad \text{s.t.} \quad  \ \delta \leq e_t(\cdot) \leq 1 - \delta, \quad   
     \ \mathsf{V} \left( e_{1}(\cdot), \dots, e_{T}(\cdot) \right) + \sum_{x \in \mathcal{X}} p(x)  \big( \tau(x) - \tau \big)^2 \leq C.\nonumber
\end{align}
In this formulation, the objective function minimizes the total expected number of failures during the entire experiment. The term inside the summation accounts for the expected failures when assigning treatment 1 with probability $e_t(x)$ and treatment $0$ with probability $1 - e_t(x)$. Here, $\mu(x, a) $ represent the expected success probabilities under treatments $a$, given covariates $x$. The first feasibility constraints again ensure that the treatment assignment probabilities remain within the range $[\delta, 1 - \delta]$. This design balances the ethical imperative to reduce adverse outcomes with statistical consideration to maintain sufficient power for detecting treatment effects.

Due to the delay mechanism's influence on the variance lower bound, the optimal treatment allocation rules differ substantially from those derived in the existing literature. We shall show that the solutions to the optimization problems in \eqref{Eq-2: Design objective 1 power maximization} and \eqref{Eq-3: Design objective 2 failure reduction}, denoted as $e_1^*(\cdot), \ldots, e_T^*(\cdot)$, form a sequence of probabilities that vary across different stages. In comparison, the classical Neyman allocation rule for power maximization:
\begin{align}\label{Eq-4: Neyman allocation}
    e_t^{\mathtt{Neyman}}(a|x)= \frac{\sigma(x,a)}{\sigma(x,1) + \sigma(x,0)}, \quad \text{for } t = 1, \ldots, T,
\end{align}
and the failure reduction rule of \cite{rosenberger2001optimal}:
\begin{align}\label{Eq-5: failure reduction rule without delay}
    e_t^{\mathtt{FR}}(a|x) = \frac{\sqrt{\mu(x,a)}}{ \sqrt{\mu(x,1)} + \sqrt{\mu(x,0)}}, \quad \text{for } t = 1, \ldots, T,
\end{align}
are both fixed probabilities that do not differ across stages. Both allocation methods ignore the delay mechanism. Therefore, when response delay plays a significant role, assigning treatments based on existing optimal rules no longer guarantees achieving the desired experimental goals.

\begin{figure}[!ht]
    \centering
\includegraphics[width=\linewidth]{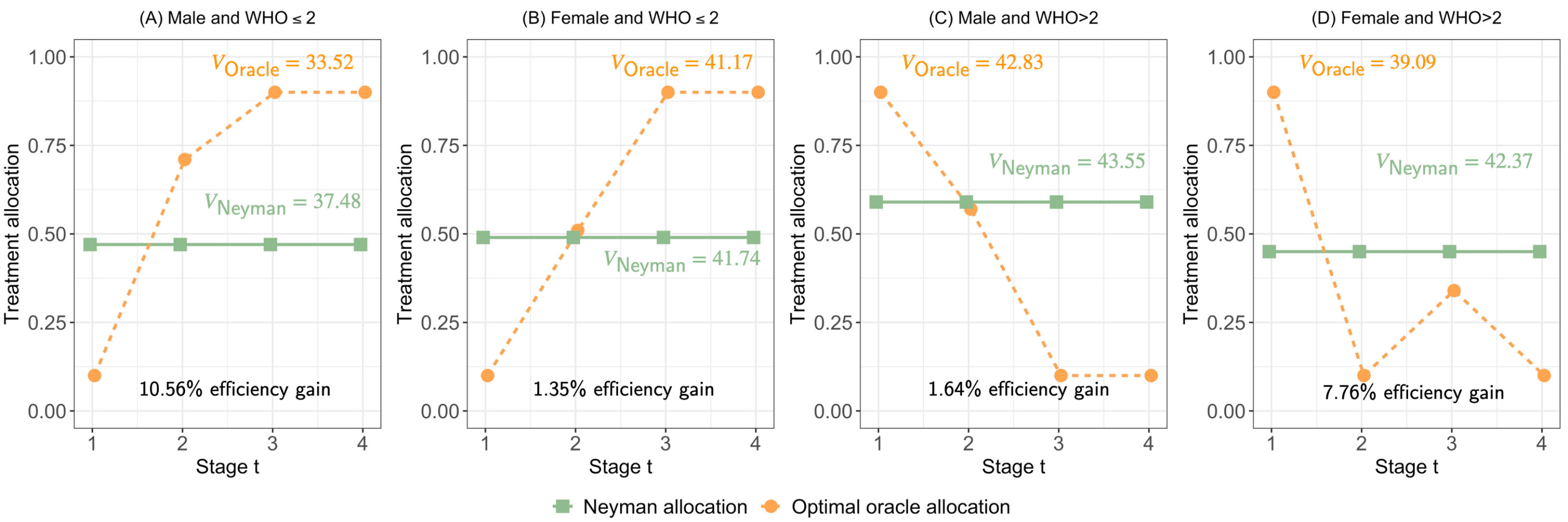}
\caption{Optimal oracle allocation (accounting for outcome delays) and the Neyman allocation for the power maximization objective. 
}
\label{fig-2:oracle design comparison X = sex, WHO stage}
\end{figure}

We illustrate this point more concretely by computing the optimal oracle treatment allocation in the example introduced in Section \ref{Section-1: Motivation and contribution}, and compare it with classical treatment assignment methods such as the Neyman allocation. The covariate strata we consider are formed by two covariates, the biological sex and the WHO clinical stages of HIV. See Section \ref{Section-5: Synthetic case study} for additional details on the model parameters. In Figure \ref{fig-2:oracle design comparison X = sex, WHO stage}, we present the oracle allocations for each stage and each covariate stratum corresponding to the power maximization objective defined in \eqref{Eq-2: Design objective 1 power maximization}, as well as the Neyman allocation. Note that, since the classical Neyman allocation ignores the delayed outcome issue, it results in a constant treatment probability across the experimental stages, whereas the oracle allocation varies significantly across both stages and covariate strata. This highlights the importance of allowing the delay mechanism to be both arm- and covariate-specific.

We also report the objective function value for each design method in the figure. As can be seen, the efficiency gain in the oracle allocation can be substantial: more than 10\% in the first covariate stratum and about 8\% in the last case. Once again, this numerical exercise demonstrates the potential gains from leveraging the delay mechanism in a CARA framework.

\section{Fully Forward-looking CARA with Delayed Outcomes}\label{Section-3: Fully forward-looking CARA with delayed outcomes}

The results provided in the previous section assume prior knowledge of the delay mechanism, and therefore, the discussed optimal treatment allocation corresponds to an infeasible oracle setting. In practice, the delay mechanism is rarely known and must be estimated as data are sequentially accumulated to optimize the experimental objective. In this section, we propose two delay-informed practical CARA designs tailored to two previously discussed experimental goals: improving statistical power \citep{tymofyeyev2007implementing} and minimizing the expected number of failures while meeting a minimum power requirement for binary responses \citep{rosenberger2001optimal}. We also provide tools for valid statistical inference after the experiment is finished. 

As with any CARA design, we use a pre-specified and fixed treatment assignment rule for the first stage due to the lack of a priori knowledge about the covariate and outcome distributions as well as the delay mechanism. In subsequent stages, our proposed fully forward-looking CARA design incorporates two key components: (i) estimating aspects of the delay mechanism from accrued data, while employing extrapolation to assess the tail behavior of the delay distribution; and (ii) using these estimates to formulate a feasible, fully forward-looking optimization problem that determines the optimal treatment assignments for the next stage. 

We first discuss the estimation of the delay mechanism. Assuming stage $t$ of the experiment has concluded, we first update our delay distribution estimator via
\begin{align}
\label{Eq: delay distribution estimation}
    \hat{\rho}_t(d|x,a) =  \sum_{\ell = 0}^d \hat{\Prob}[D_i=\ell | X_i=x,A_i=a] = \sum_{\ell = 0}^d\frac{\sum\limits_{i=1}^{N} \Indicator_{(X_{i} = x,A_{i}=a, D_{i}= \ell,S_i\leq t-\ell)}   }{\sum\limits_{i=1}^{N} \Indicator_{(X_{i} = x,A_{i}=a, S_i\leq t-\ell)}    }, \quad d\leq t-1,
\end{align}
for $a = 0,1$. Notice that the above estimator is only defined for $d\leq t-1$. To gain some intuition, consider a concrete example for $t=3$. At the end of the third stage, data collected from participants enrolled in the first stage will be informative about the delay mechanism $\rho(0|\cdot)$, $\rho(1|\cdot)$, and $\rho(2|\cdot)$. Among those whose outcome information is still missing at the end of stage 3, it is not possible to tell whether such information will become available until the end of stage 4. Following a similar reasoning, participants enrolled in the second stage can help estimate $\rho(0|\cdot)$ and $\rho(1|\cdot)$. The main message is that at any stage $t$, the delay mechanism is only estimable up to $d\leq t-1$, and the remaining ``tail features'' are not estimable until more data is collected. 

To anticipate these future delay distributions, we propose the following approaches that reflect the experimenter's expectations about future delays. For a \textit{conservative} experimenter, we consider a simple extrapolation approach: 
\begin{align}
\label{Eq: conservative extrapolation}
     \hat{\rho}_{t}(d|x,a) = \hat{\rho}_{t}(t-1|x,a) ,  \quad d \geq t.
\end{align}
This extrapolation essentially assumes that all future delay probabilities will remain at the last estimated value. As we discuss below in Section \ref{section: Convergence of the optimized treatment allocation}, the conservative extrapolation can be understood from a minimax perspective, as we will be sequentially optimizing the objective function under the worst-case scenario of the delay mechanism. For this reason, we adopt this approach in most of our numerical experiments and theoretical investigations.

For an \textit{optimistic} experimenter, we consider an extrapolation that postulates all missing outcomes will be available after one more experimental stage: $\hat{\rho}_{t}(d|x,a) = 1 , d \geq t$.
Lastly, a \textit{neutral} experimenter might gradually adjust the estimated delay probabilities from the last observed value to immediate response by interpolating between $\hat{\rho}_{t}(t-1|x,a)$ and $1$ for $d = t, \dots, T-1$. These strategies enable experimenters to incorporate future delay probabilities into our design, accommodating different perspectives on how delays may evolve. Although it is generally impossible to rank the three extrapolation methods, we show through Monte Carlo experiments that they all deliver valid inference for the desired causal effects and more effectively achieves the design objectives compared to classical approaches (such as complete randomization).

Next, to optimize treatment allocation for stage $t+1$ while accounting for response delays, we introduce a fully forward-looking estimation of the variance term $\mathsf{V}(e_{1}(\cdot), \dots, e_{T}(\cdot))$ at the current stage $t$. Our proposed estimate incorporates past allocations up to Stage $t$, future allocations from stages $t+1$ to $T$, and the estimated delay mechanisms. Our method is designed to anticipate the impact of future allocations and outcome delays on the variance of the final treatment effect estimate after the experiment concludes: 
\begin{align*}
\hat{\mathsf{V}}_t(e_{t+1}(\cdot),\dots,e_{T}(\cdot)) &=  \sum_{x\in\mathcal{X}}\hat{p}_{t}(x) \Bigg[\frac{ \hat{\sigma}_{t}^2(x,1) }
     { \sum\limits_{\ell=1}^{t}\hat{r}_\ell\hat{\rho}_{t}(T-\ell|x,1)\hat{e}^*_\ell(1|x)+ \sum\limits_{\ell=t+1}^{T}\hat{r}_\ell\hat{\rho}_{t}(T-\ell|x,1)e_{\ell}(x)} \\
   &\qquad\qquad   +\frac{ \hat{\sigma}_{t}^2(x,0) }
     {\sum\limits_{\ell=1}^{t}\hat{r}_\ell\hat{\rho}_{t}(T-\ell|x,0)\hat{e}^*_\ell(0|x)+ \sum\limits_{\ell=t+1}^{T}\hat{r}_\ell\hat{\rho}_{t}(T-\ell|x,0) (1-e_{\ell}(x))}\Bigg].
\end{align*}
Here, $\hat{p}_{t}(x) = N_t^{-1}\sum\limits_{i=1}^{N} \Indicator_{(X_{i} = x, S_i\leq t)}$, which is the estimated proportion of participants with covariate $x$ up to stage $t$. The term $\hat{r}_\ell = n_\ell / N$ denotes the proportion of participants enrolled at stage $\ell$. The estimated conditional mean and variance of the outcome $Y_i(a)$ are
\begin{align}
\label{Eq: interim estimates}
\hat{\sigma}_{t}^2(x,a) = \frac{\sum\limits_{i=1}^{N} \Indicator_{(X_{i} = x,A_{i}=a, D_{i}+S_i\leq t)} (Y_{i}-\hat{\mu}_{t}(x,a))^2  }{\sum\limits_{i=1}^{N} \Indicator_{(X_{i} = x,A_{i}=a, D_{i}+S_i\leq t)}  }, \ \hat{\mu}_{t}(x,a) = \frac{ \sum\limits_{i=1}^{N} \Indicator_{(X_{i}=x,A_{i}=a,D_{i}+S_i\leq t)}Y_{i}}{ \sum\limits_{i=1}^{N} \Indicator_{(X_{i}=x,A_{i}=a,D_{i}+S_i\leq t)}}. 
\end{align}
While the expressions may seem complicated, they are simply the sample mean and sample variance of the outcome, restricted to the subsample for group $x$ receiving treatment $a$. We remark that while the true underlying conditional means $\mu(x,a)$ and conditional variances $\sigma^2(x,a)$ do not depend on a specific stage $t$, their estimate are stage-specific. This is due to the sequential nature of the experiment: we always update previous estimates whenever new data become available. 

Using this fully forward-looking variance estimate, we can then determine an effective treatment assignment for stage $t+1$ that accounts for how future treatment allocations and response delays from stages $t+1$ onward affect the design objectives at the end of the experiment. For the first design objective of maximizing power, we solve the following optimization problem to find the treatment allocation probability for stage $t+1$:
\begin{align}\label{Eq-6: feasible, power maximization}
  (\tilde{e}_{t+1}(\cdot),\dots,\tilde{e}_{T}(\cdot)) = \argmin_{e_{t+1}(\cdot),\dots,e_{T}(\cdot)}  \ \hat{\mathsf{V}}_t(e_{t+1}(\cdot),\dots,e_{T}(\cdot)),
\end{align}
subject to the constraint that $\delta \leq e_{\ell}(\cdot)\leq 1-\delta$ for $\ell \geq t+1$. 
We then set $\hat{e}^*_{t+1}(1|\cdot) = 1 - \hat{e}^*_{t+1}(0|\cdot)=\tilde{e}_{t+1}(\cdot)$. 
 
For the second design objective of reducing failures under a power constraint, we define a similar fully forward-looking estimate of the expected proportion of total failures as
\begin{align*}
    \hat{\mathsf{P}}_t\left( e_{t+1}(\cdot), \dots, e_{T}(\cdot) \right) = & \sum_{\ell =1}^t \hat{r}_\ell \sum_{x\in\mathcal{X}} \hat{p}_t(x)\Big[ \hat{e}^*_{\ell}(1 | x) \left(1 - \hat{\mu}_t(x, 1)\right) +  \hat{e}^*_{\ell}(0| x) \left(1 - \hat{\mu}_t(x, 0)\right) \Big]   \\
    &\quad + \sum_{\ell=t+1}^T \hat{r}_\ell \sum_{x\in\mathcal{X}}\hat{p}_t(x)\Big[ e_{\ell}(x) \left(1 - \hat{\mu}_t(x, 1)\right) +  (1-e_{\ell}(x)) \left(1 - \hat{\mu}_t(x, 0)\right) \Big].
\end{align*}
We denote the solution to the following optimization problem as $(\tilde{e}_{t+1}(\cdot), \dots, \tilde{e}_{T}(\cdot))$:
\begin{align}
  \nonumber\min_{e_{t+1}(\cdot), \dots, e_{T}(\cdot)}  & \ \hat{\mathsf{P}}_t\left( e_{t+1}(\cdot), \dots, e_{T}(\cdot) \right)  \\
    \label{Eq-7: feasible, failure reduction}&\text{s.t. } 
     \hat{\mathsf{V}}_t\left( e_{t+1}(\cdot), \dots, e_{T}(\cdot) \right) +  \sum_{x \in \mathcal{X}} \hat{p}_t(x)  \big( \hat{\tau}_t(x) -  \hat{\tau}_t\big)^2 \leq C, 
\end{align}
where $\hat{\tau}_t(x) = \hat{\mu}_t(x, 1) - \hat{\mu}_t(x, 0)$ and $\hat{\tau}_t = \sum_{x\in\mathcal{X}} \hat{p}_t(x) \hat{\tau}_t(x)$. We then again set $\hat{e}^*_{t+1}(1|\cdot) = 1 - \hat{e}^*_{t+1}(0|\cdot) =  \tilde{e}_{t+1}(\cdot)$.

As both optimization problems in \eqref{Eq-6: feasible, power maximization} and \eqref{Eq-7: feasible, failure reduction} are fully forward-looking, our proposed design strategies aim to proactively adjust the allocation strategies in future stages, ensuring that the data collected accounts for the future impacts due to outcome delays. This approach contrasts with classical CARA designs that optimize trial objectives solely using historical information. At the end of the last stage $T$, the average treatment effect estimator is constructed with 
\begin{align*}
\hat{\tau}_T = \sum_{x\in\mathcal{X}} \hat{p}_T(x) \hat{\tau}_T(x)  = \sum_{x\in\mathcal{X}} \hat{p}_T(x)\Big( \hat{\mu}_T(x, 1) - \hat{\mu}_T(x, 0)\Big).
\end{align*}
We then provide a variance estimator, thus enabling valid statistical inference:
\begin{align*}
\hat{\mathsf{V}}_{T} &= \sum_{x\in \mathcal{X}} \hat{p}_T(x) \Bigg[\frac{\hat{\sigma}_T^2(x,1)}{\hat{\mathsf{e}}_T(x,1)} + \frac{\hat{\sigma}_T^2(x,0)}{\hat{\mathsf{e}}_T(x,0)} + \big(\hat{\tau}_T(x) - \hat{\tau}\big)^2\Bigg],\ 
\hat{\mathsf{e}}_T(x,a) = \frac{ \sum_{i=1}^{N} \Indicator_{(X_{i}=x,A_{i}=a,D_{i}+S_i\leq T)}}{ \sum_{i=1}^{N} \Indicator_{(X_{i}=x)}}.
\end{align*}
We summarize the analysis protocol below, where $z_{1-\frac{\alpha}{2}}$ denotes the $1-\frac{\alpha}{2}$ percentile of the standard normal distribution.
\begin{center}
\begin{tabular}{l}
\hline  \hline  
\textbf{Protocol for interim analysis and statistical inference}\hskip 4.8cm \  \\ \hline 
\textbf{Initial stage 1}: \\
\hskip 1em Enroll $n_1$ participants, and assign treatments with $\hat{e}_{1}^*(1|x)= \frac{1}{2}$.\\[2pt]
\textbf{Interim analysis after stage $t$ has concluded}: \\ 
\hskip 1em Available data: $(S_i,X_i,A_i,D_i,Y_i)$ for $i=1,2,\dots,N_t$;\\
\hskip 1em Following \eqref{Eq: delay distribution estimation} and \eqref{Eq: conservative extrapolation}, construct the delay distribution estimate $\hat{\rho}_t(d|x,a)$; \\
\hskip 1em Following \eqref{Eq: interim estimates}, construct the outcome distribution estimates $\hat{\sigma}_t(x,a)$ and $\hat{\mu}_t(x,a)$;\\
\hskip 1em From either \eqref{Eq-6: feasible, power maximization} or \eqref{Eq-7: feasible, failure reduction}, solve the allocation for stage $t+1$: $\hat{e}_{t+1}^*(1|x)$. \\[2pt]
\textbf{Treatment assignment for stage $t+1$}: \\
\hskip 1em Enroll $n_{t+1}$ participants, and assign treatments with $\hat{e}_{t+1}^*(1|x)$.\\[2pt]
\textbf{Statistical inference after the final stage $T$ has concluded}: \\
\hskip 1em Available data: $(S_i,X_i,A_i,D_i,Y_i)$ for $i=1,2,\dots,N$;\\
\hskip 1em Construct the treatment effect estimator $\hat{\tau}_T$ and the variance estimator $\hat{\mathsf{V}}_{T}$; \\
\hskip 1em Report the $(1-\alpha)\%$ confidence interval: $\hat{\tau}_T \pm z_{1-\frac{\alpha}{2}} \sqrt{\hat{\mathsf{V}}_{T}/N}$. \\[2pt] \hline\hline  
\end{tabular}
\end{center}

\section{Theoretical Investigation}\label{Section-4: Theoretical investigation}

This section provides theoretical justifications for the proposed CARA design. To begin, we present the semiparametric efficiency bound for treatment effect estimation in a multi-stage setting with delayed outcomes. We then develop a general result on the consistency and asymptotic normality of the estimated average treatment effect. Our theoretical investigation then proceeds by examining the specific setting considered in Section \ref{Section-3: Fully forward-looking CARA with delayed outcomes}, where we extrapolate the estimated delay mechanism in each stage using a conservative rule. Finally, we observe that in an experiment with only finitely many stages, it is generally not possible to achieve the oracle allocation, as this would require perfect ex ante knowledge of the delay mechanism. Therefore, the last part of our theoretical investigation addresses the efficiency loss resulting from extrapolating the delay mechanism in our design method. 

\subsection{Notation, Assumptions and the Efficiency Bound}\label{Section-4-1: Notation, assumptions and the efficiency bound}

We begin by recalling our adaptive experiment setting: the experiment consists of $T$ stages, and in each stage $t=1,2,\dots,T$, $n_t$ participants are enrolled. Let $N_t = \sum_{s=1}^t n_s$ represent the cumulative sample size at the end of stage $t$, with $N=N_T$ being the total sample size. We use the notation $S_i = t$ to indicate that individual $i$ is enrolled in stage $t$. For each individual, we observe some baseline covariates $X_{i}\in\mathcal{X}$. The observed outcome variable is denoted by $Y_{i}$, and it is related to the potential outcomes through $Y_i = A_iY_i(1) + (1-A_i)Y_i(0)$, where $A_i\in\mathcal{A}$ denotes the actual treatment assigned and $\mathcal{A} = \{0,1\}$. Moments of the potential outcomes are $\mu_s(x,a) = \Expectation[Y_i(a)^s|X_i=x]$ and $\mu_s(a) = \Expectation[Y_i(a)^s]$; we omit the subscript whenever $s=1$. 

Our first assumption imposes standard regularity conditions on the covariates and the potential outcomes.

\begin{assumption}[Covariates and potential outcomes]\label{Assu-1: Covariates and potential outcomes}\ \\
(i) The covariates and potential outcomes, $(X_i,Y_i(0),Y_i(1))$, are independent and identically distributed across $i=1,2,\dots,N$. \\
(ii) The covariates have finite support (i.e., $|\mathcal{X}|<\infty$). Let $p(x) = \Prob[X_i = x]$, then $\min_{x\in\mathcal{X}}p(x) > 0$.\\
(iii) The potential outcomes have finite fourth moments: $\max_{x\in \mathcal{X}}\max_{a\in \mathcal{A}}\mu_4(x,a) < \infty$.
\end{assumption}

The next assumption addresses the delay mechanism, where we adopt the notation $D_i$ to indicate the number of stages after which the experimenter can observe $Y_i$. By our definition, $Y_i$ is observed at the end of stage $t$ if and only if $D_i + S_i \leq t$. Also recall that the delay distribution is represented by $\rho(d|x,a) = \Prob[D_{i}  \leq d| X_{i}=x, A_{i}=a]$. 

\begin{assumption}[Delay mechanism]\label{Assu-2: Treatment assignment and delay mechanism}\ \\
(i) Outcome delay is independent of the potential outcomes after conditioning on the covariates and the treatment assignment: $D_{i} \Independence \{Y_{i}(a):a\in\mathcal{A}\}\ |\ X_{i}, A_{i}, S_i$. \\
(ii) $\min_{x\in\mathcal{X}}\min_{a\in\mathcal{A}}\rho(0|x,a) > 0$.
\end{assumption}

Finally, we consider multi-stage experiment settings.

\begin{assumption}[Asymptotic regime]\label{Assu-3: Asymptotic regime}\ \\ 
The total number of stages, $T$, is fixed, and for all $t$, $n_t/N \to r_t > 0$.
\end{assumption}

We are now ready to present the semiparametric efficiency bound for average treatment effect estimation in the oracle setting, where one has perfect knowledge of the delay mechanism and treatments are randomly assigned conditional on the covariates. This efficiency bound serves as the foundation and starting point for our proposed adaptive experimental design, as our algorithm minimizes a feasible version of the efficiency bound  constructed using accrued information.

\begin{theorem}[Semiparametric efficiency bound]\label{Thm-1: Semi-parametric efficiency bound} \ \\
Let Assumptions \ref{Assu-1: Covariates and potential outcomes}--\ref{Assu-3: Asymptotic regime} hold. In addition, assume (i) $S_i$ are independent and identically distributed with $\Prob[S_i=t] = r_t$; (ii) the treatment assignment probabilities, $e_t(a|x)=\Prob[A_i=a|X_i=x,S_i=t]$, are bounded away from 0 and 1. Then the efficient influence function for estimating $\mu(a)$ is
\begin{align*}
     \psi(X_i,A_i,D_i,Y_i,S_i|a) &= \frac{\Indicator_{(A_i=a,D_i+S_i\leq T)}}{\sum\limits_{t=1}^T r_t \rho(T-t|X_i,a)e_t(a|X_i)}\Big(Y_i(a)-\mu(X_i,a) \Big)
      + \mu(X_i,a) - \mu(a),
\end{align*}
where $a\in\mathcal{A} = \{0,1\}$. In addition, the efficient influence function for estimating $\tau$ is \begin{align*}
\psi(X_i,A_i,D_i,Y_i,S_i)\ =\ \psi(X_i,A_i,D_i,Y_i,S_i|1)-\psi(X_i,A_i,D_i,Y_i,S_i|0),
\end{align*}
leading to the semiparametric efficiency bound: 
\begin{align*}
\mathsf{V} = \sum_{x\in \mathcal{X}} p(x) \Bigg[\frac{\sigma^2(x,1)}{\sum\limits_{t=1}^T r_t  \rho(T-t|x,1)e_t(1|x)} &+ \frac{\sigma^2(x,0)}{\sum\limits_{t=1}^T r_t  \rho(T-t|x,0)e_t(0|x)}+ \big(\tau(x) - \tau\big)^2\Bigg].
\end{align*} 
\end{theorem}

We note that while the previous theorem is developed in the context of discrete covariates, it can be generalized to accommodate continuously distributed covariates (c.f., equation \ref{Eq-1: semiparametric efficiency bound}).

\subsection{Asymptotic Properties of Treatment Effects Estimates}\label{Section-4-2: Asymptotic properties of treatment effects estimates}

In this subsection, we demonstrate that the treatment effect estimator is consistent and admits an asymptotic normal distribution. The main conclusions of this section rely on a high-level condition regarding the convergence of the optimized treatment allocation rule, making these conclusions applicable to a wide range of CARA and a broad class of frequentist adaptive experimental design settings \citep{hu2004asymptotic,baldi2015almost,hu2015unified}. We begin with this high-level condition, which will be verified in the next subsection for our proposed forward-looking CARA design. 

\begin{condition}[Convergence of the optimized treatment allocation]\label{Cond-1: Convergence of the optimized treatment allocation}\ \\
There exists nonrandom $e_t(\cdot|\cdot)$, such that $\hat{e}^*_t(a|x) = e_t(a|x) + \op(1)$ for all $a\in\mathcal{A}$ and all $x\in\mathcal{X}$. 
\end{condition}

The next theorem characterizes the asymptotic normality of both the estimated subgroup treatment effects and the estimated average treatment effect. 

\begin{theorem}[Asymptotic normality of estimated treatment effects]\label{Thm-2: Asymptotic normality of estimated treatment effects}\ \\
Let Assumptions \ref{Assu-1: Covariates and potential outcomes}--\ref{Assu-3: Asymptotic regime} hold. Then under Condition \ref{Cond-1: Convergence of the optimized treatment allocation},
\begin{align*}
&\sqrt{N}\big( \hat{\tau}_T(x) - \tau(x) \big) 
\toDist \mathcal{N}\Big(0,  \frac{1}{p(x)}\Big(\frac{\sigma^2(x,1)}{\sum_{t=1}^T r_t  \rho(T-t|x,1)e_t(1|x)}+\frac{\sigma^2(x,0)}{\sum_{t=1}^T r_t  \rho(T-t|x,0)e_t(0|x)}\Big) \Big),
\end{align*}
and $\sqrt{N}\big(\hat{\tau}_T - \tau\big) \toDist \mathcal{N}(0, \mathsf{V})$, where $\mathsf{V}$ is defined in Theorem \ref{Thm-1: Semi-parametric efficiency bound}. 
\end{theorem}

It is helpful to compare this theorem to the semiparametric efficiency bound established in the previous subsection. While the two results, Theorems \ref{Thm-1: Semi-parametric efficiency bound} and \ref{Thm-2: Asymptotic normality of estimated treatment effects}, yield the same variance formula, they are conceptually very different. The semiparametric efficiency bound is derived under a fixed treatment assignment regime, which we then use to guide our design algorithms---either by minimizing the efficiency bound for power improvement or by maximizing failure reduction subject to a variance upper bound. However, from an ex ante perspective, it may not be immediately clear why this is a useful exercise, as it is not obvious how the bound could actually be achieved. Encouragingly, Theorem \ref{Thm-2: Asymptotic normality of estimated treatment effects} helps close the loop by showing that the asymptotic variance of the estimated treatment effect matches the bound in Theorems \ref{Thm-1: Semi-parametric efficiency bound} under mild regularity conditions, thereby justifying our design objective. That said, it is worth clarifying that, in general, one cannot achieve the optimized oracle efficiency bound, an issue we discuss further below. 

\begin{theorem}[Statistical inference]\label{Thm-3: Statistical inference}\ \\
Let Assumptions \ref{Assu-1: Covariates and potential outcomes}--\ref{Assu-3: Asymptotic regime} hold. Then under Condition \ref{Cond-1: Convergence of the optimized treatment allocation}, $\hat{\mathsf{V}}_{T} = {\mathsf{V}}_{T} + \op(1)$.
\end{theorem}

\subsection{Convergence of the Optimized treatment Allocation}\label{section: Convergence of the optimized treatment allocation}\label{Section-4-3: Convergence of the optimized treatment allocation}

Previously we have shown that the treatment effect estimators are consistent and asymptotically normally distributed. These results build on the high-level condition that the empirical treatment allocation, $\hat{e}_t^*(\cdot)$, will converge in large samples (Condition \ref{Cond-1: Convergence of the optimized treatment allocation}). In this subsection, we further investigate the theoretical properties of the proposed experimental design, specifically, solved from \eqref{Eq-6: feasible, power maximization}, and show that it satisfies Condition \ref{Cond-1: Convergence of the optimized treatment allocation}, thereby tying all loose ends. To give a road map, we will first show that the asymptotic analogs of our design algorithm have unique solutions. In this process, we will carefully distinguish the oracle problem of minimizing the semiparametric efficiency bound (which requires perfect knowledge of the delay mechanism) from the problem of sequential optimization using only accrued knowledge (which requires extrapolating the estimated delay mechanism). Next, we show that the empirically optimized treatment allocation converges to some large-sample limit. To conclude this subsection, we will present a bound on efficiency loss due to extrapolating the delay mechanism. 

In our theoretical investigation, we will focus on the conservative method for estimating the delay mechanism. This allows us to provide simple and easy-to-interpret conditions, hence avoiding lengthy discussions. Another appealing property of the conservative approach is that it can be understood from a ``minimax'' perspective, where we sequentially minimize the asymptotic variance under the worst-case scenario of the delay mechanism. Despite our focus on one particular approach in this subsection, we show in our simulations that all three methods (conservative, optimistic, and neutral) deliver valid inference. 

To start, we recall that $\mathsf{V}(e_1(\cdot),\dots,e_T(\cdot))$ is the objective function in \eqref{Eq-2: Design objective 1 power maximization}. Its minimizers will be denoted by $e_t^*(a|x)$ for $a\in\{0,1\}$, $x\in\mathcal{X}$, and $t=1,2,\dots,T$. As we have discussed, the oracle optimal solution is not achievable in general, as it builds on perfect knowledge of the delay mechanism. In any realistic experimental setting, however, this information can only be learned sequentially. In fact, even we set $n_t = \infty$, part of the delay mechanism can still remain unknown for any $t$: to be even more precise, one typically only learn $\rho(d|x,a)$ for $d<t$ using information at the end of stage $t$. For this reason, we have introduced feasible and concrete ways to extrapolate the delay mechanism. Following our discussion in Section \ref{Section-2: Optimal delay-adjusted treatment allocation in CARA}, we define $\rho^\dagger_t(d|x,a) = \min\big\{\rho(d|x,a), \rho(t-1|x,a)\big\}.$

It is helpful to compare this definition with Assumption \ref{Assu-2: Treatment assignment and delay mechanism}. There, we assumed that the delay mechanism is time-homogeneous, so that $\rho(d|x,a)$ is not indexed by $t$. On the other hand, $\rho^\dagger_t(d|x,a)$ arises due to our imperfect knowledge about the delay mechanism and the conservative extrapolation employed, and as a result it will depend on a specific stage. It is not the true delay mechanism, but it is the asymptotic analogue of $\hat{\rho}_t(d|x,a)$, which we used in our optimization algorithm.  
To study convergence property of $\hat{e}^*_t(a|x)$, we will make a definition first. Let $e^\dagger_1(0|x) = e^\dagger_1(1|x) = 1/2$. Then for $t=1,2,\dots,T$, define recursively that
\begin{align*}
e^\dagger_{t+1}(1|\cdot) = 1 - e^\dagger_{t+1}(0|\cdot) = \argmin_{e_{t+1}(\cdot)} \min_{e_{t+2}(\cdot),\dots,e_{T}(\cdot)}\mathsf{V}_t^\dagger(e_{t+1}(\cdot),\dots,e_{T}(\cdot)),
\end{align*}
where 
\begin{align*}
    \mathsf{V}_t^\dagger(e_{t+1}(\cdot),\dots,e_T(\cdot)) &= \sum_{x\in\mathcal{X}}p(x)\Bigg[\frac{\sigma^2(x,1)}{\sum\limits_{\ell=1}^{t}r_\ell \rho^\dagger_t(T-\ell|x,1)e_\ell^\dagger(1|x) +  \sum\limits_{\ell=t+1}^{T}r_\ell \rho^\dagger_t(T-\ell|x,1)e_{\ell}(x)} \\
    & +  \frac{\sigma^2(x,0)}{\sum\limits_{\ell=1}^{t}r_\ell \rho^\dagger_t(T-\ell|x,0)e_\ell^\dagger(0|x) +  \sum\limits_{\ell=t+1}^{T}r_\ell \rho^\dagger_t(T-\ell|x,0)(1-e_\ell(x))}\Bigg].
\end{align*}

Our first main result is to show that both $e_t^*(a|x)$ and $e_t^\dagger(a|x)$ are unique. To this end, we make the following assumption. 

\begin{assumption}[Variation in delay mechanism]\label{Assu-4: Variation in delay mechanism}\ \\
For any $d\neq d'$, ${\rho(d|x,1)}/{ \rho(d|x,0)} \neq {\rho(d'|x,1)}/{ \rho(d'|x,0)}$.
\end{assumption}

We remark that this assumptions helps establish the uniqueness of $e_t^*(a|x)$. However, it is still possible to have ${\rho_t^\dagger(d|x,1)}/{ \rho_t^\dagger(d|x,0)} = {\rho_t^\dagger(d'|x,1)}/{ \rho_t^\dagger(d'|x,0)}$. Fortunately, this is easily resolved by a tie-breaking rule. See the Supplementary Materials for a specific recommendation and additional discussions. 

\begin{theorem}[Consistent treatment allocation]\label{Thm-4: Consistent treatment allocation}\ \\
Let Assumptions \ref{Assu-1: Covariates and potential outcomes}--\ref{Assu-4: Variation in delay mechanism} hold. Then for $a\in\{0,1\}$, $x\in\mathcal{X}$, and $t=1,2,\dots,T$, both $e_t^*(a|x)$ and $e_t^\dagger(a|x)$ are unique. In addition, Condition \ref{Cond-1: Convergence of the optimized treatment allocation} holds with $\hat{e}_t^*(a|x) \toProb e_t^\dagger(a|x) $.
\end{theorem}
  
Before closing this section, we provide two insights regarding the variance minimization problem. The first result suggests that as more information about the delay mechanism becomes available in the adaptive experiment, the optimized asymptotic variance always (weakly) decreases. While this result may seem natural, we note that it relies on the use of conservative extrapolation. As discussed earlier, conservative extrapolation can be understood from a minimax perspective, where the asymptotic variance is sequentially minimized under the worst-case scenario of the delay mechanism.

The second result provides a bound on efficiency loss due to imperfect knowledge about the delay mechanism. Collectively, these two results provide theoretical guarantee to our proposed adaptive experimental design.

\begin{theorem}[Optimized variance and bound on efficiency loss]\label{Thm-5: Optimized variance and bound on efficiency loss}\ \\
Let Assumptions \ref{Assu-1: Covariates and potential outcomes}--\ref{Assu-4: Variation in delay mechanism}  hold. Define $\mathsf{V}^*$ as the minimized oracle asymptotic variance, and $\mathsf{V}$ to be the asymptotic variance of the estimated average treatment effect. Then (i) for all $1\leq s< t\leq T-1$, 
\begin{align*}
&\mathsf{V}^*
\ \leq\ 
\mathsf{V}
\ \leq\ 
\min_{e_{t+1}(\cdot),\dots,e_T(\cdot)}\mathsf{V}_t^\dagger(e_{t+1}(\cdot),\dots,e_T(\cdot))
\ \leq\ 
\min_{e_{s+1}(\cdot),\dots,e_T(\cdot)}\mathsf{V}_s^\dagger(e_{s+1}(\cdot),\dots,e_T(\cdot));
\end{align*}
and (ii) for any $\epsilon > 0$ and $d_\epsilon$ with $\min_{x\in\mathcal{X},a\in\mathcal{A}}\rho(d_{\epsilon}|x,a) \geq 1-\epsilon$, $\mathsf{V} \leq \mathsf{V}^* + C(d_{\epsilon}T^{-1} + \epsilon)$, where $C$ does not depend on $T$, $\epsilon$ or $d_\epsilon$.
\end{theorem}

\section{Synthetic Case Study}\label{Section-5: Synthetic case study}

To evaluate the performance of the proposed forward-looking CARA design, we conduct simulation studies using a data generation mechanism calibrated from the study of \cite{fahey2020financial}. For the first set of simulation results, we consider biological sex as the only covariate $X_{i}$, leading to two subgroups: $X_{i}=\mathtt{M}$ and $X_{i}=\mathtt{F}$ with frequencies 36\% and 64\%, respectively. We set  $n_t =100$ for $t=1, \ldots,T$ with $T\in\{4,6,8\}$. We report simulation evidence separately for the two design objectives in \eqref{Eq-2: Design objective 1 power maximization} and \eqref{Eq-3: Design objective 2 failure reduction}, which require different transformation of the outcome variable.\bigskip

\noindent \textbf{Setup 1: Power maximization.} For the power maximization design objective, we compare four design methods: (i) our proposed design; (ii) our design enhanced by the doubly adaptive biased coin design (DBCD) \citep{hu2004asymptotic,tymofyeyev2007implementing}; (iii) complete randomization; (iv) Neyman allocation. We note that DBCD is a response-adaptive randomization design that adjusts treatment allocation toward the optimal allocation dynamically. Since our approach involves assigning treatments across multiple covariate strata, we apply the DBCD design separately within each stratum. We also follow the convention in the literature and use the natural logarithm of the viral load as the outcome variable. Following the trial data, the conditional mean and standard deviation of the potential outcomes are $\mu(\mathtt{F},1)=2.50$, $\mu(\mathtt{F},0)=2.98$, $\sigma(\mathtt{F},1)=0.36$, $\sigma(\mathtt{F},0)=2.06$ for the female subgroup, and $\mu(\mathtt{M},1)=2.47$, $\mu(\mathtt{M},0)=2.72$, $\sigma(\mathtt{M},1)=0.82$, $\sigma(\mathtt{M},0)=0.31$ for the male subgroup. Given the calibrated model primitives, the average treatment effect is $\tau = -0.33$. To evaluate the power of different designs under the power maximization objective, we also vary the magnitude of the ATE in our data-generating process by perturbing $\mu(\mathtt{M},0)$, such that the perturbed ATE takes values between $0$ and $0.66$. \bigskip

\noindent \textbf{Setup 2: Failure reduction.}  For the failure reduction objective, we again compare across four designs: (i--ii) the proposed method (with and without DBCD enhancement); (iii) the ethical design \citep{rosenberger2001optimal}; (iv) complete randomization. In this simulation setup, we transform the original viral load measurements into a binary variable, indicating whether the participant has achieved viral suppression; specifically, if a participant's viral load is below $1,000$ copies per mL, we set $Y_i = 1$ (viral suppression). Conditional means of the potential outcomes are $\mu(\mathtt{F},1)=0.78$, $\mu(\mathtt{F},0)=0.57$, $\mu(\mathtt{M},1)=0.84$ and $\mu(\mathtt{M},0)=0.63$. The average treatment effect is $\tau = 0.21$. To evaluate the power of different designs under the failure reduction objective, we vary the magnitude of ATE in our data-generating process by perturbing $\mu(\mathtt{M},0)$, such that the perturbed ATE takes values between  $0$ and $0.42$. 

As we discussed in Section \ref{Section-1: Motivation and contribution} and illustrated in Figure \ref{Fig-1: empirical example, delay distribution, X = sex}, the outcomes in the original data exhibit both arm- and covariate-dependent delays, which we also incorporate in our simulation setup. To be precise, the delay variable is generated conditionally according a multinomial distribution: $D_{i}|(X_{i}=x, A_{i}=a)\sim \text{Multinomial}\left(1,\newline \bm{\rho}(x,a)\right)$. Here, we slightly abuse the notation to define $\bm{\rho}(x,a) = (\mathbb{P}(D_i = 0|x,a),\ldots, \mathbb{P}(D_i = T-1|x,a))$, where each element represents the probability of delaying by exactly $d$ stages (instead of cumulative). For $T=4$, the delay mechanism parameters are $\bm{\rho}(\texttt{F},1) = (0.64,0.18,0.07,0.03)$ and $\bm{\rho}(\texttt{F},0) = (0.63,0.18,0.05,0.02)$ for the female subgroup, and $\bm{\rho}(\texttt{M},1) = (0.55,0.23,0.10,0.02)$ and $\bm{\rho}(\texttt{M},0) = (0.54,$ $0.11, 0.21,  0.01)$ for the male subgroup. 

We show in panels (A)--(C) of Figure \ref{fig-3:simulation X=sex only} the optimized variance under the power maximization objective. For complete randomization, the variances of the treatment effect estimator are 7.75, 8.04, and 8.17, respectively. The results clearly demonstrate that the proposed fully forward-looking CARA design and its DBCD-enhanced version exhibit a smaller deviation from the oracle variance (indicated by the horizontal black dashed line); to compare, complete randomization leads to severe efficiency loss. Interestingly, the Neyman allocation performs reasonably well in this design, although it still leads to slightly larger variances. 

\begin{figure}[!ht]
\centering
\includegraphics[width=1\linewidth]{Revision_figures/main-fig-3-1-simulated-var.pdf}
\includegraphics[width=1\linewidth]{Revision_figures/main-fig-3-2-simulated-failure.pdf}
\caption{Optimized variance (panels A--C) and failure rate (panels D--F) for experimental horizons $T = 4, 5, 6$. Covariate stratification is based on biological sex. 
}
\label{fig-3:simulation X=sex only}
\end{figure}

Panels (D)--(F) in the figure collects the failure rates for both our proposed method and the ethical design. For complete randomization, the failure rates are 0.31,0.33, and 0.35. Again, our proposed method and its DBCD-enhanced version tend to perform better, and the overall failure rate there is very close to the oracle level (horizontal black dashed line). In contrast, failure rates tend to be considerably higher for complete randomization. 

To assess the finite-sample distributional properties of the treatment effect estimator, we present the coverage probabilities in Table \ref{table-1:coverage probability X=sex} for the two design objectives and the three extrapolations methods for the estimated delay mechanism. Encouragingly, the empirical coverage is extremely close to the nominal level. Lastly, we demonstrate and compare the power properties of different design strategies in Figure \ref{fig-4:simulation:power curve}: all methods control the type I error rate well when the true ATE is zero; with $\tau > 0$, however, our CARA design has a clear advantage as it is able to detect the non-zero treatment effect more frequently. 

In summary, this synthetic case study showcases that our fully forward-looking CARA design not only provides valid statistical inference in the presence of delayed responses, but it also delivers higher estimation efficiency and enhanced failure reduction for the two design objectives.

\begin{table}[!ht]
\centering
\caption{Coverage probability comparison. }
\label{table-1:coverage probability X=sex}
\resizebox{0.85\textwidth}{!}{ 
\begin{tabular}{lcccc}
\hline\hline
\multicolumn{5}{c}{\textbf{Design 1: Power Maximization}} \\
\cline{2-5}
Delay perspective & \text{Proposed CARA design} & \text{DBCD + Proposed CARA design} &  &  \\
\midrule
\text{Conservative} & 0.94~(0.01) & 0.95~(0.01)  &  & \\
\text{Optimistic} & 0.95~(0.01) & 0.94~(0.01)  &  & \\
\text{Neutral} & 0.95~(0.01) & 0.95~(0.01)  & & \\
\hline
\multicolumn{5}{c}{\textbf{Design 2: Failure Reduction}} \\
\cline{2-5}
Delay perspective & \text{Proposed CARA design} & \text{DBCD + Proposed CARA design} & & \\
\midrule
\text{Conservative} & 0.94~(0.01) & 0.95~(0.01) &  &  \\
\text{Optimistic} & 0.96~(0.01) & 0.94~(0.01) &  &  \\
\text{Neutral} & 0.95~(0.01) & 0.94~(0.01) &  &  \\
\hline\hline
\end{tabular}
}
\end{table}

\begin{figure}[!ht]
\centering
\includegraphics[width=0.67\linewidth]{Revision_figures/main-fig-4-power.pdf}
\caption{Power comparison for the two design objectives. Covariate stratification is based on biological sex. 
}
\label{fig-4:simulation:power curve}
\end{figure}

Before closing this section, we provide additional simulation evidence where stratification is based on biological sex and the WHO clinical stages of HIV. Specifically, we define four groups: $\texttt{S}_1$ for male subjects with WHO stages 1 and 2, $\texttt{S}_2$ for female subjects with stages 1 and 2, $\texttt{S}_3$ for male with stages 3 and 4, and $\texttt{S}_4$ for female with stages 3 and 4. 

For the first stratum $\texttt{S}_1$, the calibrated parameters are 
$\sigma(\texttt{S}_1,1)=1.63$,  
$\sigma(\texttt{S}_1,0)=1.87$, 
$\bm{\rho}(\texttt{S}_1,1) =(0.10, 0.15, 0.32,0.04)$, and 
$\bm{\rho}(\texttt{S}_1,0) =(0.03,0.10,0.22,0.30)$. 
For the second stratum $\texttt{S}_2$, the parameters are 
$\sigma(\texttt{S}_2,1)=1.85$, 
$\sigma(\texttt{S}_2,0)=1.90$, 
$\bm{\rho}(\texttt{S}_2,1) =(0.14,0.16,0.21,0.05)$ and 
$\bm{\rho}(\texttt{S}_2,0) =(0.10,0.14,0.17,0.06)$. 
For $\texttt{S}_3$, we adopt 
$\sigma(\texttt{S}_3,1)=2.37$, 
$\sigma(\texttt{S}_3,0)=1.63$,
$\bm{\rho}(\texttt{S}_3,1) =(0.10,0.19,0.21,0.14)$ and 
$\bm{\rho}(\texttt{S}_3,0) =(0.11,0.16,0.19,0.09)$. 
And finally for $\texttt{S}_4$, we use 
$\sigma(\texttt{S}_4,1)=1.73$, 
$\sigma(\texttt{S}_4,0)=2.14$,
$\bm{\rho}(\texttt{S}_4,1) =(0.05,0.20,0.17,0.18)$ and 
$\bm{\rho}(\texttt{S}_4,0) =(0.19,0.08,0.24,0.02)$. The oracle treatment allocations under the proposed design and the Neyman allocation are demonstrated earlier in Figure \ref{fig-2:oracle design comparison X = sex, WHO stage}. 

We present simulation results on the optimized variances in panels (A)–(C) of Figure \ref{fig-5:simulation variance X=sex WHO stage}, and the failure rates in (D)--(F). Although the covariate stratification differs (cf. Figure \ref{fig-3:simulation X=sex only}), a consistent pattern emerges: our proposed design yields lower variance and enhanced failure reduction by explicitly accounting for the delay mechanism. To compare, the variances of the treatment effect estimator employing complete randomization are 43.96, 45.17, and 47.82, and the failure rates are 0.34, 0.32, and 0.34, respectively. To complement our earlier findings in Figure \ref{fig-4:simulation:power curve}, we also report power comparisons in Figure \ref{fig-6:simulation:power curve X = sex WHO stage}. 

\begin{figure}[!ht]
\centering
\includegraphics[width=1\linewidth]{Revision_figures/main-fig-5-1-simulated-var-gender-stage.pdf}
\includegraphics[width=1\linewidth]{Revision_figures/main-fig-5-2-simulated-failure-gender-stage.pdf}
\caption{Optimized variance (panels A--C) and failure rate (panels D--F) for experimental horizons $T = 4, 5, 6$. Covariate stratification is based on biological sex and WHO clinical stages. 
}
\label{fig-5:simulation variance X=sex WHO stage}
\end{figure}

\begin{figure}[!ht]
    \centering
\includegraphics[width=0.67\linewidth]{Revision_figures/main-fig-6-power-gender-stage.pdf}
\caption{Power comparison for the two design objectives. Covariate stratification is based on biological sex and WHO clinical stages. 
}
\label{fig-6:simulation:power curve X = sex WHO stage}
\end{figure}

\section{Conclusion}

In this paper, we introduce fully forward-looking covariate-adjusted response-adaptive (CARA) designs that effectively address the challenge of delayed outcomes in adaptive experiments, allowing the delay mechanism to be both arm- and covariate-dependent. Our approach sequentially estimates the delay mechanism, which is then used to inform optimal treatment allocation. Through a comprehensive synthetic case study, we demonstrate that our proposed design can lead to substantial gains in both experimental efficiency and participant welfare compared to traditional CARA designs. We provide rigorous theoretical foundation by deriving semiparametric efficiency bounds in the presence of delayed responses and establishing the consistency and asymptotic normality of treatment effect estimators.

\section*{Supplementary Material}
The Supplementary Material \citep{supp-ref} contains general theoretical results, their proofs, and reports additional simulation evidence.

\section*{Acknowledgments}
We thank the editor, the associate editor, two anonymous reviewers, seminar participants, and conference audiences for their valuable feedback, which significantly improved the manuscript. Xinwei Ma gratefully acknowledges financial support from the National Institutes of Health through the grant R01-AG089512. Jingshen Wang gratefully acknowledges financial support from the National Institutes of Health through the grant R01-AG089512, and the National Science Foundation through the grants DMS-2239047 and DMS-2220537.

\bibliography{reference}
\bibliographystyle{jasa}
\end{document}


\begin{center}
\Large Supplementary Material to \\
``Covariate-Adjusted Response-Adaptive Design with Delayed Outcomes''
\end{center}

\vskip 3em

\begin{center}
\begin{tabular}{ccc}
Xinwei Ma & Jingshen Wang & Waverly Wei \\ 
&& \\
\footnotesize Department of Economics & \footnotesize Division of Biostatistics & \footnotesize Department of Data Sciences  \\
\footnotesize  & \footnotesize  & \footnotesize  and Operations \\
\footnotesize University of California San Diego & \footnotesize University of California Berkeley & \footnotesize University of Southern California
\end{tabular}
\end{center}

\vskip 2em

\begin{center}
\footnotesize \today \\
\end{center}

\vskip 3em
\onehalfspacing

{\footnotesize \tableofcontents}

\vfill

\newpage

\onehalfspacing

\setcounter{page}{1}
\pagestyle{plain}

\section{Setup, Notation and Assumptions} 

We begin by recalling our adaptive experiment setting: There are $T$ stages, which are labeled by $t=1,2,\dots,T$. In each stage $t$, $n_t$ participants are enrolled. Participants are labeled by $i=1,2,\dots,N$, with $N = \sum_{t=1}^T n_t$ being the total sample size. We employ the notation, $S_i = t$ to indicate that the $i$th individual is enrolled in stage $t$. 

For each individual, we observe some baseline covariates, $X_{i}\in\mathcal{X}$. The observed outcome variable is denoted by $Y_{i}$. As individuals are randomized into control and treatment arms, we use $A_{i}\in\mathcal{A} = \{0,1\}$ to represent the actual treatment status. Lastly, since the outcome information may not be immediately available after the treatment is assigned, we use $D_{i}$ to represent the number of stages after which the experimenter can observe $Y_{i}$. As a result of our definition, $Y_i$ is observed at the end of stage $t$ if and only if $D_i + S_i \leq t$. Following the Neyman-Rubin framework, the potential outcomes are denoted by $Y_{i}(a)$ for different treatments $a\in\mathcal{A} = \{0,1\}$. 

In what follows, we collect some notation and the main assumptions adopted in this supplementary material as well as in the main paper. 

\begin{assumption}\label{assu-1: Covariates and potential outcomes}\ \\
(i) The covariates and potential outcomes, $\{X_{i}, \{Y_{i}(a):a\in\mathcal{A}\}\}$, are independent and identically distributed across $i=1,2,\dots,N$. \ \\
(ii) The covariates have a finite support, that is, $|\mathcal{X}|<\infty$. Let
\begin{align*}
p(x) = \Prob[X_i = x],
\end{align*}
then $\min_{x\in\mathcal{X}}p(x) > 0$.\ \\
(iii) Let 
\begin{align*}
\mu_s(x,a) = \Expectation\big[Y_i(a)^s|X_i=x\big].
\end{align*}
Then $\max_{x\in \mathcal{X}}\max_{a\in \mathcal{A}}\mu_4(x,a) < \infty$.\qed
\end{assumption}
For future reference, we also define the unconditional moments of the potential outcomes as
\begin{align*}
\mu_s(a) = \Expectation\big[Y_i(a)^s\big].
\end{align*}
Conditional and unconditional average treatment effects are
\begin{align*}
\tau(x) = \mu_1(x,1) - \mu_1(x,0),\quad \tau = \mu_1(1) - \mu_1(0).
\end{align*}
Conditional variances of the potential outcomes are denoted by
\begin{align*}
    \sigma^2(x,a) &= \Var[Y_{i}(a)|X_i=x,A_i=a] = \mu_2(x,a) - \mu_1(x,a)^2.
\end{align*}

We remark that in the main paper, we drop the subscript whenever $s=1$; that is, for the mean of the potential outcomes, we use
\begin{align*}
\mu(x,a) = \mu_1(x,a) = \Expectation\big[Y_i(a)|X_i=x\big]\text{ and }\mu(a) =  \mu_1(a) = \Expectation\big[Y_i(a)\big].
\end{align*}

The next assumption concerns the treatment assignment and the delay mechanism. Recall that $\mathcal{H}_t$ represents the history up to time $t$, and $\mathcal{H}_0$ is the trivial sigma-algebra. 

\begin{assumption}\label{assu-2: Treatment assignment and delay mechanism}\ \\
(i) Conditioning on $S_{i}=t$, $X_i$, and $\mathcal{H}_{t-1}$, treatment $A_i$ is randomly assigned, and
\begin{align*}
\Prob[A_{i} = a|X_{i}=x,S_i=t,\mathcal{H}_{t-1}] = \hat{e}^*_t(a|x),
\end{align*}
where $\hat{e}^*_t(\cdot|x)\in \mathcal{H}_{t-1}$. \ \\
(ii) The delay mechanism is  independent of the potential outcomes after conditioning on the covariates and the treatment assignment: 
\begin{align*}
D_{i} \Independence \{Y_{i}(a):a\in\mathcal{A}\}\ |\ X_{i}, A_{i}, S_i.
\end{align*}
In addition, the delay mechanism is time-homogeneous:
\begin{align*}
\rho(d|x,a) = \Prob[D_{i}  \leq d| X_{i}=x, A_{i}=a] = \Prob[D_{i}  \leq d| X_{i}=x, A_{i}=a, S_i=t] . 
\end{align*}
\vskip-2em\qed
\end{assumption}

We also assume the treatment assignment probability is bounded away from zero, and that there is a strictly positive probability that the outcome information can be immediately observed. 

\begin{assumption}\label{assu-3: Overlap}\ \\
(i) With probability 1, $\min_{x\in\mathcal{X}}\min_{a\in\mathcal{A}}\min_{t\leq T}\hat{e}^*_t(a|x) \geq \delta$ for some $\delta > 0$. \ \\
(ii) $\min_{x\in\mathcal{X}}\min_{a\in\mathcal{A}}\rho(0|x,a) > 0$.\qed
\end{assumption}

Finally, we consider an asymptotic regime in which $T$ is fixed, and $n_t\to \infty$. More precisely, we assume
\begin{assumption}\label{assu-4: Asymptotic regime}\ \\
The total number of stages, $T$, is fixed, and for all $t$, 
\begin{align*}
    \frac{n_t}{N} \to r_t > 0.
\end{align*}
Here, $N = \sum_{t=1}^T n_t$ is the total sample size. \qed
\end{assumption}

\section{Semiparametric Efficiency Bound}

In this section, we establish the semiparametric efficiency bound for treatment effect estimation in the presence of delayed outcomes. 

\begin{thm}\label{SA-thm-1: Semiparametric efficiency bound}\ \\
Let Assumptions \ref{assu-1: Covariates and potential outcomes}--\ref{assu-4: Asymptotic regime} hold. In addition, assume (i) $S_i$ are independently and identically distributed with $\Prob[S_i=t] = r_t$; (ii) $\hat{e}_t^*(a|x) = {e}_t(a|x)$ for some ${e}_t(\cdot|\cdot)$ that  does not depend on $\mathcal{H}_{t-1}$. Then the efficient influence function for estimating $\mu_1(a)$ is
\begin{align*}
\psi(X_i,A_i,D_i,Y_i,S_i|a) &= \frac{\Indicator_{(A_i=a,D_i+S_i\leq T)}}{\sum\limits_{t=1}^T r_t e_t(a|X_i)\rho(T-t|X_i,a)}\Big(Y_i(a)-\mu_1(X_i,a) \Big)
+ \mu_1(X_i,a) - \mu_1(a).
\end{align*}
In addition, the efficient influence function for estimating $\tau$ is
\begin{align*}
&\psi(X_i,A_i,D_i,Y_i,S_i)\ =\ \psi(X_i,A_i,D_i,Y_i,S_i|1)-\psi(X_i,A_i,D_i,Y_i,S_i|0)\\[8pt]
=&\frac{\Indicator_{(A_i=1,D_i+S_i\leq T)}}{\sum\limits_{t=1}^T r_t e_t(1|X_i)\rho(T-t|X_i,1)}\Big(Y_i(1)-\mu_1(X_i,1) \Big)\\
&\qquad\qquad -\frac{\Indicator_{(A_i=0,D_i+S_i\leq T)}}{\sum\limits_{t=1}^T r_t e_t(0|X_i)\rho(T-t|X_i,0)}\Big(Y_i(0)-\mu_1(X_i,0) \Big) + \tau(X_i) - \tau.
\end{align*}
As a result, the semiparametric efficiency bound for estimating $\tau$ is 
\begin{align*}
\mathsf{V} = \sum_{x\in \mathcal{X}} p(x) \Bigg[\frac{\sigma^2(x,1)}{\sum\limits_{t=1}^T r_t e_t(1|x) \rho(T-t|x,1)} &+ \frac{\sigma^2(x,0)}{\sum\limits_{t=1}^T r_t e_t(0|x) \rho(T-t|x,0)}+ \big(\tau(x) - \tau\big)^2\Bigg].
\end{align*} 
\vskip-2em\qed
\end{thm}

Condition (ii) assumes that the treatment allocation rule, ${e}_t$, is not random. That is, we treat ${e}_t$ as ``pre-determined'' or ``fixed'' in our semiparametric efficiency calculation. Given the semiparametric efficiency bound, we develop our adaptive experimental design algorithm to minimize the asymptotic variance of the treatment effect estimators. 

\section{Treatment Effect Estimation and Statistical Inference}

In this section, we show that the main treatment effect estimator is consistent following our proposed adaptive experimental design, and admits an asymptotically normal distribution. We also provide a valid variance estimator. Main conclusions of this section build on a ``high-level'' condition on the consistency of the optimized treatment allocation rule. We verify this high-level condition in the next subsection. 

To start, we recall that the estimated mean potential outcome for a specific subgroup is
\begin{align*}
\hat{\mu}_{s,T}(x,a) &= \frac{ \sum_{i=1}^{N} \Indicator_{(X_{i}=x,A_{i}=a,D_{i}+S_i\leq T)}Y_{i}^s}{ \sum_{i=1}^{N} \Indicator_{(X_{i}=x,A_{i}=a,D_{i}+S_i\leq T)}\phantom{Y_{i}^s}}.
\end{align*}
Notice that outcome information may not be observed for all observations. In our notation above, this is captured by the indicator function $\Indicator_{(D_{i}+S_i\leq T)}$, which equals 1 if and only if outcome information for this individual is available at the end of the experiment. 

Whenever necessary, we also use the notation $\hat{\mu}_{s,t}(x,a)$ to represent the estimator formed with observations in the first $t$ stages; that is, 
\begin{align*}
\hat{\mu}_{s,t}(x,a) &= \frac{ \sum_{i=1}^{N} \Indicator_{(X_{i}=x,A_{i}=a,D_{i}+S_i\leq t)}Y_{i}^s}{ \sum_{i=1}^{N} \Indicator_{(X_{i}=x,A_{i}=a,D_{i}+S_i\leq t)}{\color{white}Y_{i}^s}}.
\end{align*}

Consistency of the estimator follows from the lemma below. 

\begin{lem}\label{SA-lem-1: Consistent estimation of mean potential outcomes}\ \\
Let Assumptions \ref{assu-1: Covariates and potential outcomes}--\ref{assu-4: Asymptotic regime} hold. Then for all $s\leq 2$, 
\begin{align*}
\hat{\mu}_{s,T}(x,a) &= \mu_{s}(x,a) + \Op\left(\sqrt{\frac{1}{N}}\right).
\end{align*}
\vskip-2em\qed
\end{lem}

Next, we show that $\hat{\mu}_{s,T}(x,a)$ admits an asymptotic normal distribution. To this end, we employ the following high-level condition, which we verify for our proposed design in the next section. 

\begin{condition}\label{cond-1: Consistent treatment allocation}\ \\
There exists nonrandom $e_t(\cdot|\cdot)$, such that $\hat{e}^*_t(a|x) = e_t(\cdot|\cdot) + \op(1)$ for all $a\in\mathcal{A}$ and all $x\in\mathcal{X}$. \qed
\end{condition}

\begin{lem}\label{SA-lem-2: Asymptotic normality of estimated mean potential outcomes}\ \\
Let Assumptions \ref{assu-1: Covariates and potential outcomes}--\ref{assu-4: Asymptotic regime} hold. Then under Condition \ref{cond-1: Consistent treatment allocation},
\begin{align*}
&\sqrt{N}\big( \hat{\mu}_{1,T}(x,a) - \mu_1(x,a) \big) \\
&\quad = \frac{1}{p(x)\sum_{t=1}^T r_t e_t(a|x) \rho(T-t|x,a)}\frac{1}{\sqrt{N}} \sum_{i=1}^{N} \Indicator_{(X_{i}=x,A_{i}=a,D_{i}+S_i\leq T)}\big(Y_{i} - \mu_1(x,a)\big) + \op(1) \\[5pt]
&\quad \toDist \mathcal{N}\Big(0,  \frac{\sigma^2(x,a)}{p(x)\sum_{t=1}^T r_t e_t(a|x) \rho(T-t|x,a)} \Big)
\end{align*}
for $a\in\mathcal{A} = \{0,1\}$.
\qed
\end{lem}

Define estimated subgroup treatment effect:
\begin{align*}
\hat{\tau}_T(x) &= \hat{\mu}_{1,T}(x,1) - \hat{\mu}_{1,T}(x,0),
\end{align*}
then it is immediate from Lemma \ref{SA-lem-2: Asymptotic normality of estimated mean potential outcomes} that $\hat{\tau}_T(x)$ also admits an asymptotic normal distribution. 

\begin{thm}\label{SA-thm-2: Asymptotic normality of estimated subgroup treatment effects}\ \\
Let Assumptions \ref{assu-1: Covariates and potential outcomes}--\ref{assu-4: Asymptotic regime} hold. Then under Condition \ref{cond-1: Consistent treatment allocation},
\begin{align*}
&\sqrt{N}\big( \hat{\tau}_T(x) - \tau(x) \big) \toDist \mathcal{N}\Big(0,  \frac{1}{p(x)}\Big(\frac{\sigma^2(x,1)}{\sum_{t=1}^T r_t e_t(1|x) \rho(T-t|x,1)}+\frac{\sigma^2(x,0)}{\sum_{t=1}^T r_t e_t(0|x) \rho(T-t|x,0)}\Big) \Big).
\end{align*}
\vskip-2em\qed
\end{thm}

Finally, we consider the estimated treatment effect: 
\begin{align*}
\hat{\tau}_T &= \sum_{x\in \mathcal{X} } \hat{p}_T(x)\hat{\tau}_T(x),
\end{align*}
where 
\begin{align*}
\hat{p}_T(x) = \frac{1}{N} \sum_{i=1}^{N} \Indicator_{(X_{i}=x)}. 
\end{align*}

\begin{thm}\label{SA-thm-3: Asymptotic normality of estimated treatment effects}\ \\
Let Assumptions \ref{assu-1: Covariates and potential outcomes}--\ref{assu-4: Asymptotic regime} hold. Then under Condition \ref{cond-1: Consistent treatment allocation},
\begin{align*}
\sqrt{N}\big(\hat{\tau}_T - \tau\big) \toDist \mathcal{N}(0, \mathsf{V}),
\end{align*}
where
\begin{align*}
\mathsf{V} = \sum_{x\in \mathcal{X}} p(x) \Bigg[\frac{\sigma^2(x,1)}{\sum_{t=1}^T r_t e_t(1|x) \rho(T-t|x,1)} &+ \frac{\sigma^2(x,0)}{\sum_{t=1}^T r_t e_t(0|x) \rho(T-t|x,0)}  + \big(\tau(x) - \tau\big)^2\Bigg].
\end{align*}
\vskip-2em\qed
\end{thm}

Following the asymptotic normality result established in the above theorem, we now introduce a variance estimator. Define
\begin{align*}
    \hat{\sigma}_{T}(x,a) = \hat{\mu}_{2,T}(x,a) - \hat{\mu}_{1,T}(x,a)^2
\end{align*}
for $x\in \mathcal{X}$ and $a\in \mathcal{A}$. Next, let
\begin{align*}
\hat{\mathsf{e}}_T(x,a) &= \frac{ \sum_{i=1}^{N} \Indicator_{(X_{i}=x,A_{i}=a,D_{i}+S_i\leq T)}}{ \sum_{i=1}^{N} \Indicator_{(X_{i}=x)}}.
\end{align*}
Finally, we define
\begin{align*}
\hat{\mathsf{V}}_{T} &= \sum_{x\in \mathcal{X}} \hat{p}_T(x) \Bigg[\frac{\hat{\sigma}_T^2(x,1)}{\hat{\mathsf{e}}_T(x,1)} + \frac{\hat{\sigma}_T^2(x,0)}{\hat{\mathsf{e}}_T(x,0)} + \big(\hat{\tau}_T(x) - \hat{\tau}_T\big)^2\Bigg].
\end{align*}
The result below establishes valid statistical inference. 

\begin{thm}\label{SA-thm-4: Statistical inference}\ \\
Let Assumptions \ref{assu-1: Covariates and potential outcomes}--\ref{assu-4: Asymptotic regime} hold. Then under Condition \ref{cond-1: Consistent treatment allocation},
\begin{align*}
\hat{\mathsf{V}}_{T} = {\mathsf{V}}_{T} + \op(1).
\end{align*}
Therefore, 
\begin{align*}
\frac{\hat{\tau}_T - \tau}{\sqrt{\hat{\mathsf{V}}_{T} /N}} \toDist \mathcal{N}(0,1).
\end{align*}
\vskip-2em\qed
\end{thm}

\section{Numerical Algorithm}

Minimizing the objective function can be challenging, especially when $T$ is large. As part of our methodological development, we first discuss a dimension reduction technique that maps the original optimization into a one dimensional convex problem, thereby greatly allowing fast and scalable implementation. To convey the main idea, we first consider the oracle problem, for which the underlying data-generating process (the potential outcomes distribution and the delay mechanism) is assumed to be known. However, we remark that the same algorithm can be applied to compute the feasible allocations in our fully forward-looking method, to be discussed below.
To present the algorithm in its full generality, consider the following objective function
\begin{align*}
\mathsf{V}(e_1,e_2,\dots,e_J) &:= \frac{\mathsf{c}_1}{\mathsf{a}_0 + \mathsf{a}_1 e_1 + \dots + \mathsf{a}_J e_J} + \frac{\mathsf{c}_0}{\mathsf{b}_0 + \mathsf{b}_1 (1-e_1) + \dots + \mathsf{b}_J (1-e_J)},
\end{align*}
with the constraints that $e_j\in[\delta,1-\delta]$ for some pre-specified $\delta\in[0,1/2)$. To gain some intuition, we can easily map the above objective function to the semiparametric efficiency bound, as the example below illustrates. 

\begin{example}\ \\
We will fix a specific covariate value $x$. Set $J=T$, and
\begin{align*}
\mathsf{c}_1 = \sigma^2(x,1),\ \mathsf{c}_0 = \sigma^2(x,0),\ \mathsf{a}_j=r_j \rho(T-j|x,1),\ \mathsf{b}_j=r_j \rho(T-j|x,0). 
\end{align*}
Then the optimization problem can be recast as $\min_{e_1,e_2,\dots,e_T}\mathsf{V}(e_1,\dots,e_T)$. We also introduce two additional constants $\mathsf{a}_0$ and $\mathsf{b}_0$ into the objective function to accommodate a pilot stage in which the treatment allocation is fixed (say, to 0.5 or $1/|\mathcal{A}|$). However, this is not required: our algorithm proposed below allows $\mathsf{a}_0 = \mathsf{b}_0 = 0$.\qed
\end{example}

\begin{algorithm}[!ht]
\renewcommand{\arraystretch}{2.5}
\caption{Optimal treatment allocation with dimension reduction  \\ 
\phantom{aaaaaaaaaaaa..}See Section \ref{section: R code} for \textsf{R} implementation}\label{algo:dim reduction}
\begin{algorithmic}[1]
\Statex \hskip -0.55cm\textbf{Initialization}:
  \State Fix the constants $\mathsf{c}_1,\mathsf{c}_0$, $\mathsf{a}_0,\dots,\mathsf{a}_J$, $\mathsf{b}_0, \dots,\mathsf{b}_J$.
  \State Let $\ell(1),\dots,\ell(J)$ be a permutation of $1,\dots,J$, such that
\begin{align*}
\frac{\mathsf{a}_{\ell(1)}}{\mathsf{b}_{\ell(1)}} \leq \frac{\mathsf{a}_{\ell(2)}}{\mathsf{b}_{\ell(2)}} \leq \dots \leq \frac{\mathsf{a}_{\ell(J)}}{\mathsf{b}_{\ell(J)}}.
\end{align*}
\State Define a point in $\mathbb{R}^2$, $(\alpha_0, \beta_0)$, as
\begin{align*}
\alpha_0= \mathsf{a}_0+(1-\delta)\sum_{j=1}^J \mathsf{a}_{j},\ \  \beta_0 = \mathsf{b}_0 + \delta J.
\end{align*}
\Statex \hskip -0.55cm\textbf{Characterizing boundaries of $\mathcal{V}$}: 
\For{$j\text{ from}\ 1 \ \text{to}\  J$} 
\State Define a point in $\mathbb{R}^2$, $(\alpha_j, \beta_j)$, as
\begin{align*}
\alpha_j=\mathsf{a}_0  + (1-\delta)\sum_{s=j+1}^J \mathsf{a}_{\ell(s)}+ \delta j,\ \  \beta_j = \mathsf{b}_0 + (1-\delta)\sum_{s=1}^{j} \mathsf{b}_{\ell(s)} + \delta (J-j).
\end{align*}
\State Define a point in $\mathbb{R}^2$, $(\underline{\alpha}_j, \underline{\beta}_j)$, as
\begin{align*}
\underline{\alpha}_j=\mathsf{a}_0 + (1-\delta)\sum_{s=1}^{J-j} \mathsf{a}_{\ell(s)} + \delta j,\ \ \underline{\beta}_j = \mathsf{b}_0 + (1-\delta)\sum_{s=J-j+1}^{J} \mathsf{b}_{\ell(s)} + \delta (J-j).
\end{align*}
\EndFor
\State Upper boundary of $\mathcal{V}$: connect points $(\alpha_0, \beta_0)\cdots(\alpha_1, \beta_1)
\cdots(\alpha_2, \beta_2)$ $\cdots\cdots(\alpha_J, \beta_J)$. 
\State Lower boundary of $\mathcal{V}$: connect points $(\underline{\alpha}_0, \underline{\beta}_0)\cdots(\underline{\alpha}_1, \underline{\beta}_1)\cdots(\underline{\alpha}_2, \underline{\beta}_2)\cdots\cdots(\underline{\alpha}_J, \underline{\beta}_J)$. 
\Statex \hskip -0.55cm\textbf{Optimization}:
\State Solve the optimization problem, $\min_{\alpha,\beta} \mathsf{V}(\alpha,\beta)$ subject to $(\alpha,\beta)\in\mathcal{V}$. 
\State With the solution $(\alpha^{ *},\beta^*)$, find the corresponding $e_1^*,\dots,e_J^*$. \label{algo:last step}
\end{algorithmic}
\end{algorithm}

Our strategy is to first re-parameterize the problem as 
\begin{align*}
\min_{\alpha,\beta}\mathsf{V}(\alpha,\beta) &:= {\mathsf{c}_1}\big/{\alpha}\ +\ {\mathsf{c}_0}\big/{\beta},\qquad \text{subject to\ }(\alpha,\beta)\in\mathcal{V}\subset \mathbb{R}^2.
\end{align*}
The set $\mathcal{V}$ is determined by the linear mappings 
$$(e_1,\dots,e_J)\mapsto \alpha(e_1,\dots,e_J) := \mathsf{a}_0 + \mathsf{a}_1 e_1 + \dots + \mathsf{a}_Je_J,$$
and
$$(e_1,\dots,e_J)\mapsto \beta(e_1,\dots,e_J) := \mathsf{b}_0 + \mathsf{b}_1 (1-e_1) + \dots + \mathsf{b}_J (1-e_J).$$
The two denominators, $\alpha$ and $\beta$,  can also be interpreted as the probability of observing the outcome in the two treatment arms, respectively, given a specific propensity score profile (treatment assignment rule). As a key ingredient of our methodological development, Algorithm \ref{algo:dim reduction} characterizes $\mathcal{V}$ as a convex set with piecewise linear boundaries in $\mathbb{R}^2$, which makes the optimization problem fast and scalable. The main idea behind the algorithm is to compare the (estimated) delay mechanism across the two treatment arms. If it is anticipated that units in one arm tend to exhibit longer delays, then they are allocated to this arm in relatively early stages. Also see Section \ref{section: R code} for \textsf{R} implementation of the algorithm.

The last step in our algorithm finds the optimal treatment allocation from the solution $(\alpha^{*},\beta^*)\in\mathbb{R}^2$. Before discussing this step in more detail, the following result formally justifies the validity of our algorithm.

\begin{lem}\label{SA-lem-3: unique solution beta prime beta in the reduced problem}\ \\
The optimization problem of $\min_{\alpha,\beta} \mathsf{V}(\alpha,\beta)$ subject to $(\alpha,\beta)\in\mathcal{V}$ admits a unique solution, and the solution lies on the upper boundary of $\mathcal{V}$, which consists of line segments connecting points $(\alpha_0, \beta_0)\cdots(\alpha_1, \beta_1)\cdots\cdots(\alpha_J, \beta_J)$ specified in the algorithm.\qed
\end{lem}

We highlight that our algorithm and the theoretical result above apply in broader settings concerning treatment allocations in adaptive experiments. In particular, our dimension reduction technique, which maps the original problem into a univariate optimization, is applicable provided that the objective function takes a ``single-index structure'' where the allocation probabilities enter via a linear index.  

The uniqueness of the solution to the minimization of $\mathsf{V}(\alpha,\beta)$ does not imply that the original minimization problem also admits a unique solution. We further characterize the solution to the original problem. 

\begin{thm}\label{SA-thm-5: solution to the original optimization problem}\ \\
Assume the solution lies on a line segment connecting $(\alpha, \beta_j)\cdots(\alpha_{j+1}, \beta_{j+1})$, then 
\[
\frac{\mathsf{a}_{s}}{\mathsf{b}_{s}}  < \frac{\mathsf{a}_{\ell(j+1)}}{\mathsf{b}_{\ell(j+1)}}\ \Rightarrow\ e_{s}^*=\delta,\ \ \frac{\mathsf{a}_{s}}{\mathsf{b}_{s}}  > \frac{\mathsf{a}_{\ell(j+1)}}{\mathsf{b}_{\ell(j+1)}}\ \Rightarrow\ e_{s}^*=1-\delta.
\]
The solution $(e_1^*,\dots,e_T^*)$ is unique if and only if there does not exist $s$ such that $\frac{\mathsf{a}_{s}}{\mathsf{b}_{s}}  = \frac{\mathsf{a}_{\ell(j+1)}}{\mathsf{b}_{\ell(j+1)}} $.\qed
\end{thm}

In applications, the oracle treatment assignment is not feasible as it relies on unknown features of potential outcomes, such as their conditional second moments and the delay mechanism. Therefore, a feasible algorithm must estimate these features using accrued information and adjust the treatment allocation sequentially. A key challenge is that in any experimental stage, certain ``tail features'' of the delay mechanism are not estimable. To give a concrete example, assume the researcher has collected data from two stages and is revising the treatment assignment for Stage 3. Employing available data, it is possible to estimate the probability of outcome information becoming immediately available or delayed by one stage. However, it is not possible to estimate, say, the probability of delayed outcomes by three stages. This makes global optimization of the efficiency bound generally impossible. To address this challenge, we propose two feasible algorithms for allocating treatments based on different specifications of the delay mechanism. Notably, our dimension reduction technique can be readily extended to create feasible algorithms that are both numerically stable and fast. 

\subsection*{One-step forward algorithm} 

We first consider a one-step forward algorithm that uses collected experimental data to determine the optimal treatment allocation for the next one stage. To start, we initialize stage 1 of the experiment following the completely randomized design in line \ref{line:stage1-1} of Algorithm \ref{algo:feasible algorithm}. Fix some $t=1,2,\dots,T-1$, and assume we have collected data for the first $t$ stages of the experiment. We then compute the optimal treatment allocation for the next stage, $t+1$. Define the objective function
\begin{align*} 
\hat{\mathsf{V}}^{\mathtt{one-step}}_t(e_{t+1}(\cdot)) = \sum_{x\in\mathcal{X}} \hat{p}_{t}(x)\Bigg[&\frac{ \hat{\sigma}_{t}^2(x,1) }{ \sum\limits_{\ell=1}^{t}\hat{\mathsf{a}}_\ell(x)\hat{e}_\ell^*(1|x) +\hat{\mathsf{a}}_{t+1}(x) e_{t+1}(x)}\\
&\qquad\qquad\qquad+\frac{ \hat{\sigma}_{t}^2(x,0) }{\sum\limits_{\ell=1}^{t} \hat{\mathsf{b}}_{\ell}(x) \hat{e}_\ell^*(0|x)+ \hat{\mathsf{b}}_{t+1}(x)(1-e_{t+1}(x))}\Bigg].
\end{align*}
\begin{align*}
\text{with}\ \hat{p}_{t}(x) &= \frac{\sum\limits_{i=1}^{N} \Indicator_{(X_{i} = x,S_i\leq t)} }{\sum\limits_{\ell=1}^{t}n_\ell },\quad 
\hat{\sigma}_{t}^2(x,a) = \frac{\sum\limits_{i=1}^{N} \Indicator_{(X_{i} = x,A_{i}=a, D_{i}+S_i\leq t)} (Y_{i}-\hat{\mu}_{1,t}(x,a))^2  }{\sum\limits_{i=1}^{N} \Indicator_{(X_{i} = x,A_{i}=a, D_{i}+S_i\leq t)}  },
\end{align*}
where we recall that $\hat{\mu}_{1,t}(x,a)$ is simply the sample average of the outcome variable in subgroup $x$ and treatment arm $a$. Other constants in the algorithm are defined analogously:
\begin{alignat*}{3}
\hat{\mathsf{a}}_\ell(x) &= \hat{r}_\ell\hat{\rho}_{t}(t-\ell|x,1),\quad &&\hat{\mathsf{a}}_{t+1}(x) &&= \hat{r}_{t+1}\hat{\rho}_{t}(0|x,1),\\
\hat{\mathsf{b}}_\ell(x) &= \hat{r}_\ell\hat{\rho}_{t}(t-\ell|x,0),\quad
&&\hat{\mathsf{b}}_{t+1}(x) &&= \hat{r}_{t+1}\hat{\rho}_{t}(0|x,0),
\end{alignat*} 
with $\hat{r}_\ell = n_\ell / N$. 

A key challenge is to estimate the delay mechanism. At the end of stage $t$, we are able to estimate part of the delay mechanism as
\begin{align*}
\hat{\rho}_{t}(d|x,a) &= \sum_{\ell = 0}^d \hat{\Prob}[D_i=\ell | X_i=x,A_i=a] = \sum_{\ell = 0}^d\frac{\sum\limits_{i=1}^{N} \Indicator_{(X_{i} = x,A_{i}=a, D_{i}= \ell,S_i\leq t-\ell)}   }{\sum\limits_{i=1}^{N} \Indicator_{(X_{i} = x,A_{i}=a, S_i\leq t-\ell)}    },
\end{align*}
for $d=0, 1, \dots, t-1$. However, no estimate for ${\rho}(t|x,a)$ will be available until the end of stage $t+1$. Therefore, to operationalize the one-step forward objective function, we propose two approaches: (1) \texttt{conservative}: $\hat{\rho}_{t}(t|x,a) = \hat{\rho}_{t}(t-1|x,a)$, assuming that the delayed outcomes will continue to be delayed in the next stage, and (2) \texttt{optimistic}: $ \hat{\rho}_{t}(t|x,a) = 1$, which assumes all the delayed outcomes will be observed in the next stage. 

We remark that the conservative extrapolation can be understood from a ``minimax'' perspective, where we sequentially minimize the asymptotic variance under the worst-case scenario of the delay mechanism. For this reason, we adopt this approach in most of our numerical experiments, as well as in the following section of this Supplementary Material.

At the end of stage $T$ when the experiment concludes, we obtain the subgroup treatment effect estimates $\hat{\tau}_T(x)$ as well as the estimated ATE $\hat{\tau}_T$.

\begin{center}
\begin{algorithm}[!ht]
\renewcommand{\arraystretch}{1.2}
\caption{Feasible adaptive experimental design with one-step forward objective function}\label{algo:feasible algorithm}
\begin{algorithmic}[1]
\Statex \hskip -0.55cm\textbf{Initialization:}
  \State  Enroll $n_1$ participants, and assign treatments with $\hat{e}_{1}^*(1|x) = \hat{e}_{1}^*(0|x) = \frac{1}{2}$; \label{line:stage1-1}
\Statex \hskip -0.55cm \textbf{Adaptive treatment allocation:} 
\For{$t\rightarrow 1 \ \text{to}\  T-1$} 
\State Using $\mathcal{H}_{t}$, obtain $\hat{\sigma}_{t}(x,1)$, $\hat{\sigma}_{t}(x,0)$, $\hat{\rho}_{t}(d|x,1)$, and $\hat{\rho}_{t}(d|x,0)$ for $x\in \mathcal{X}$ and $d\leq t-1$. 
\State Estimate $\hat{\rho}_{t}(t|x,a)$ with either the \texttt{conservative} or the \texttt{optimistic} approach. 
\State Construct the constants $\hat{\mathsf{a}}_\ell(x)$ and $\hat{\mathsf{b}}_\ell(x)$ for $\ell = 1,2,\dots,t+1$. 
\State Minimize $\hat{\mathsf{V}}^{\mathtt{one-step}}_t$ to obtain $\hat{e}^*_{t+1}(x)$ for $x\in\mathcal{X}$.
\State Set $\hat{e}^*_{t+1}(1|x) = \hat{e}^*_{t+1}(x)$, and $\hat{e}^*_{t+1}(0|x) = 1-\hat{e}^*_{t+1}(x)$.
\State Enroll a new set of $n_{t+1}$ subjects and assign treatments. 
\EndFor
\end{algorithmic}
\end{algorithm}
\end{center}

\subsection*{Fully forward-looking algorithm} 
Our previous algorithm optimizes an objective function that involves only one additional period, which can be quite different from the semiparametric efficiency bound. While this might be innocuous when it gets closer to the last stage of the experiment, the one-step method may not perform well in the early stages. A more suitable and more sophisticated approach will require optimizing all future treatment allocations in each design stage. Specifically, we define the following fully forward-looking objective function
\begin{align*} 
&\hat{\mathsf{V}}^{\mathtt{forward}}_t(e_{t+1}(\cdot),\dots,e_{T}(\cdot)) \\
&\qquad= \sum_{x\in\mathcal{X}}\hat{p}_{t}(x) \Bigg[\frac{ \hat{\sigma}_{t}^2(x,1) }
 { \sum\limits_{\ell=1}^{t}\hat{\mathsf{a}}_\ell(x)\hat{e}_\ell^*(1|x)+ \sum\limits_{\ell=t+1}^{T}\hat{\mathsf{a}}_{\ell}(x) e_{\ell}(x)} +\frac{ \hat{\sigma}_{t}^2(x,0) }
 {\sum\limits_{\ell=1}^{t} \hat{\mathsf{b}}_{\ell}(x) \hat{e}_\ell^*(0|x)+ \sum\limits_{\ell=t+1}^{T}\hat{\mathsf{b}}_{\ell}(x) (1-e_{\ell}(x))}\Bigg].
\end{align*}

As the above algorithm is fully forward-looking at each stage, we expect it to be better aligned with the objective of variance minimization. It is also worth mentioning that our earlier Algorithm \ref{algo:dim reduction} can be employed to solve the forward-looking optimization problem: one simply needs to replace the unknown quantities in \ref{algo:dim reduction} by their estimates. It offers the same benefit of dimension reduction, which can be valuable in the early stages of the experiments. A feasible design algorithm employing the fully forward-looking algorithm is given below. 

\begin{center}
\begin{algorithm}[!ht]
\renewcommand{\arraystretch}{1.2}
\caption{Feasible adaptive experimental design with fully forward-looking objective function}\label{algo:feasible algorithm 2}
\begin{algorithmic}[1]
\Statex \hskip -0.55cm\textbf{Initialization:}
  \State  Enroll $n_1$ participants, and assign treatments with $\hat{e}_{1}^*(1|x) = \hat{e}_{1}^*(0|x) = \frac{1}{2}$; 
\Statex \hskip -0.55cm \textbf{Adaptive treatment allocation:} 
\For{$t\rightarrow 1 \ \text{to}\  T-1$} 
\State Using $\mathcal{H}_{t}$, obtain $\hat{\sigma}_{t}(x,1)$, $\hat{\sigma}_{t}(x,0)$, $\hat{\rho}_{t}(d|x,1)$, and $\hat{\rho}_{t}(d|x,0)$ for $x\in \mathcal{X}$ and $d\leq t-1$. 
\State Estimate $\hat{\rho}_{t}(d|x,a)$ for $d\geq t$ with one of the three approaches discussed below. 
\State Construct the constants $\hat{\mathsf{a}}_l(x)$ and $\hat{\mathsf{b}}_l(x)$ for $\ell = 1,2,\dots,T$. 
\State Minimize $\hat{\mathsf{V}}^{\mathtt{forward}}_t$ to obtain $\hat{e}^*_{t+1}(x),\dots,\hat{e}^*_{T}(x)$ for $x\in\mathcal{X}$.
\State Set $\hat{e}^*_{t+1}(1|x) = \hat{e}^*_{t+1}(x)$, and $\hat{e}^*_{t+1}(0|x) = 1-\hat{e}^*_{t+1}(x)$.
\State Enroll a new set of $n_{t+1}$ subjects and assign treatments. 
\EndFor
\end{algorithmic}
\end{algorithm}
\end{center}

Similar to the one-step algorithm, the delay mechanism is not fully estimable at each stage. For example, at the end of stage $t$, one can only estimate $\rho(0|x,a),\dots,\rho(t-1|x,a)$ while the fully forward-looking algorithm requires estimates for $\rho(t|x,a),\dots,\rho(T-1|x,a)$. Several candidates are: (1) \texttt{conservative}: $\hat{\rho}_{t}(d|x,a) = \hat{\rho}_{t}(t-1|x,a)$ for all $d \geq t$; (2) \texttt{optimistic}: $ \hat{\rho}_{t}(d|x,a) = 1$; and (3) \texttt{neutral}: $\hat{\rho}_{t}(d|x,a)$ linear interpolates $\hat{\rho}_{t}(t-1|x,a)$ and 1 for $d = t,\dots, T-1$.

\section{Allocation Consistency and Verification of Condition \ref{cond-1: Consistent treatment allocation}}

In this section we discuss in more details theoretical properties of the optimized treatment allocation. As part of this endeavor, we will provide sufficient conditions under which the optimized treatment allocations converge to a nonrandom limit, that is, we will verify Condition \ref{cond-1: Consistent treatment allocation}. Together with the regularity conditions laid out earlier, such design consistency helps establish both consistency and asymptotic normality of the estimated treatment effects. We will adopt the fully forward-looking algorithm with the conservative extrapolation approach.

In what follows, we will define recursively two types of treatment allocations, denoted by $e^\dagger_t(\cdot|\cdot)$ and $e^*_t(\cdot|\cdot)$, respectively. Loosely speaking, $e^\dagger_t(\cdot|\cdot)$ corresponds to the optimized treatment allocation rule if one has access to an infinite sample but imperfect knowledge about the delay mechanism, while $e^*_t(\cdot|\cdot)$ is the optimized allocation if one has both an infinite sample and perfect knowledge about the delay mechanism. To start, we set
\begin{align*}
e_1^\dagger(1|x) = e_1^*(1|x) = \frac{1}{2},\qquad e_1^\dagger(0|x) = e_1^*(0|x) = \frac{1}{2}.
\end{align*}
Now fix some $t=1,2,\dots,T-1$: the first $t$ stages of the experiment has concluded and we are optimizing treatment allocation for the next stage, $t+1$. Define
\begin{align*}
    \rho_t^\dagger(d|x,1) = \begin{cases}
        \rho(d|x,1) &\text{if }d < t\\
        \rho(t-1|x,1) &\text{if }d \geq t
    \end{cases},\qquad \rho^\dagger_t(d|x,0) = \begin{cases}
        \rho(d|x,0) &\text{if }d < t\\
        \rho(t-1|x,0) &\text{if }d \geq t
    \end{cases}.
\end{align*}
Then define
\begin{align*}
    \mathsf{V}_t^\dagger(e_{t+1}(\cdot),\dots,e_T(\cdot)) &= \sum_{x\in\mathcal{X}}p(x)\Bigg[\frac{\sigma^2(x,1)}{\sum\limits_{\ell=1}^{t}r_\ell \rho^\dagger_t(T-\ell|x,1)e_\ell^\dagger(1|x) +  \sum\limits_{\ell=t+1}^{T}r_\ell \rho^\dagger_t(T-\ell|x,1)e_\ell(x)} \\
    & +  \frac{\sigma^2(x,0)}{\sum\limits_{\ell=1}^{t}r_\ell \rho^\dagger_t(T-\ell|x,0)e_\ell^\dagger(0|x) +  \sum\limits_{\ell=t+1}^{T}r_\ell \rho^\dagger_t(T-\ell|x,0)(1-e_\ell(x))}\Bigg],
\end{align*}
and
\begin{align*}
    \mathsf{V}_t(e_{t+1}(\cdot),\dots,e_T(\cdot)) &= \sum_{x\in\mathcal{X}}p(x)\Bigg[\frac{\sigma^2(x,1)}{\sum\limits_{\ell=1}^{t}r_\ell \rho(T-\ell|x,1)e_\ell^*(1|x) +  \sum\limits_{\ell=t+1}^{T}r_\ell \rho(T-\ell|x,1)e_\ell(x)} \\
    &\qquad +  \frac{\sigma^2(x,0)}{\sum\limits_{\ell=1}^{t}r_\ell \rho(T-\ell|x,0)e_\ell^*(0|x) +  \sum\limits_{\ell=t+1}^{T}r_\ell \rho(T-\ell|x,0)(1-e_\ell(x))}\Bigg].
\end{align*}
Intuitively, $\mathsf{V}_t^\dagger$ can be understood as the large-sample analogue of the feasible objective function $\hat{\mathsf{V}}_t^{\mathtt{forward}}$, while  $\mathsf{V}_t$ is the oracle objective function that requires full knowledge of the data generating process and the delay mechanism. 

The optimized treatment allocations for the next stage $t+1$ are
\begin{align*}
    e_{t+1}^\dagger(1|x) = 1 - e_{t+1}^\dagger(0|x) = \argmin_{e_{t+1}(\cdot)}\min_{e_{t+2}(\cdot),\dots,e_T(\cdot)} \mathsf{V}^\dagger_t(e_{t+1}(\cdot),\cdots,e_T(\cdot)),
\end{align*}
and
\begin{align*}
    e_{t+1}^*(1|x) = 1 - e_{t+1}^*(0|x) = \argmin_{e_{t+1}(\cdot)}\min_{e_{t+2}(\cdot),\dots,e_T(\cdot)}  \mathsf{V}_t(e_{t+1}(\cdot),\cdots,e_T(\cdot)).
\end{align*}

The following assumption will greatly help simplify some of the presentations. We remark that this assumption can be dropped with the expense of much lengthier proof and more cumbersome notation. 

\begin{assumption}\label{assu-5: Variation in delay mechanism}\ \\
For any $d\neq d'$, 
\begin{align*}
    \frac{\rho(d|x,1)}{ \rho(d|x,0)} \neq \frac{\rho(d'|x,1)}{ \rho(d'|x,0)}.
\end{align*}
\vskip-2em\qed
\end{assumption}

We remark that it is still possible to have ``ties'' in $\rho_t^\dagger$ due to the conservative extrapolation we employed. Specifically, for $t \leq T/2$, the following always hold:
\begin{align*}
\frac{\rho_t^\dagger(T-t-1|x,1)}{\rho_t^\dagger(T-t-1|x,0)} =  \frac{\rho_t^\dagger(T-t-2|x,1)}{\rho_t^\dagger(T-t-2|x,0)} = \dots = \frac{\rho_t^\dagger(t-1|x,1)}{\rho_t^\dagger(t-1|x,0)}.
\end{align*}
Such ties will also arise when solving Algorithm \ref{algo:feasible algorithm 2}, that is, for all $t\leq T/2$, 
\begin{align*}
\frac{\hat{\mathsf{a}}_{t+1}(x)}{\hat{\mathsf{b}}_{t+1}(x)} =  \frac{\hat{\mathsf{a}}_{t+2}(x)}{\hat{\mathsf{b}}_{t+2}(x)} = \dots = \frac{\hat{\mathsf{a}}_{T-t+1}(x)}{\hat{\mathsf{b}}_{T-t+1}(x)}
\end{align*}
by construction. Therefore, we adopt the following tie-breaking rule: 

\begin{condition}\label{condition-2: Tie-breaking}\ \\ 
For all $t=1,2,\dots, T/2$, 
\begin{align*}
&\max \Big\{e^\dagger_{t+2}(1|x),\dots, e^\dagger_{T-t+1}(1|x)\Big\} > \delta \quad \text{only if}\quad e^\dagger_{t+1}(1|x) = 1-\delta,\\
&\max \Big\{\hat{e}^*_{t+2}(1|x),\dots, \hat{e}^*_{T-t+1}(1|x)\Big\} > \delta \quad \text{only if}\quad \hat{e}^*_{t+1}(1|x) = 1-\delta.
\end{align*}
\vskip-2em\qed
\end{condition}
While this condition may seem complicated, it is quite innocuous and is also straightforward to implement. Essentially, whenever the design algorithm (either the feasible allocation from minimizing the fully forward-looking objective function or its asymptotic analogue) dictates assigning some participants to the treatment arm in periods $t+1$, $t+2$, $\dots$, $T-t+1$, we will prioritize period $t+1$.  

As a direct consequence of Theorem \ref{SA-thm-5: solution to the original optimization problem}, the optimized allocations are unique. We formally state this result in the Corollary below. 

\begin{coro}\label{SA-coro:Uniqueness of optimized allocations}\ \\
Let Assumptions \ref{assu-1: Covariates and potential outcomes}--\ref{assu-5: Variation in delay mechanism} and Condition \ref{condition-2: Tie-breaking} hold. Then both $e_t^\dagger(\cdot|\cdot)$ and $e_t^*(\cdot|\cdot)$ are unique for $t=1,2,\dots,T$. \qed
\end{coro}

The next theorem is our main design consistency result, which demonstrates that the empirically optimized treatment allocation converges in large samples. 

\begin{thm}\label{SA-thm-6: Consistent treatment allocation}\ \\
Let Assumptions \ref{assu-1: Covariates and potential outcomes}--\ref{assu-5: Variation in delay mechanism} and Condition \ref{condition-2: Tie-breaking} hold. Then Condition \ref{cond-1: Consistent treatment allocation} holds with $e_t(\cdot|\cdot) =e_t^\dagger(\cdot|\cdot) $. \qed
\end{thm}

Before closing this section, we provide two insights regarding the variance minimization problem. The first result suggests that as more information about the delay mechanism becomes available in the adaptive experiment, the optimized asymptotic variance never increases. While this result may seem natural, we note that it relies on the use of conservative extrapolation. As discussed earlier, conservative extrapolation can be understood from a ``minimax'' perspective, where the asymptotic variance is sequentially minimized under the worst-case scenario of the delay mechanism.

The second result provides a bound on efficiency loss due to imperfect knowledge about the delay mechanism. Collectively, these two results provide theoretical guarantee to our proposed adaptive experimental design. 

\begin{thm}\label{SA-thm-7: Optimized variance and bound on efficiency loss}\ \\ 
Let Assumptions \ref{assu-1: Covariates and potential outcomes}--\ref{assu-5: Variation in delay mechanism} and Condition \ref{condition-2: Tie-breaking} hold. 
Define $\mathsf{V}^*$ as the minimized oracle asymptotic variance, and $\mathsf{V}$ to be the asymptotic variance of the estimated average treatment effect. Then (i) for all $1\leq s< t\leq T-1$, 
\begin{align*}
&\mathsf{V}^*
\ \leq\ 
\mathsf{V}
\ \leq\ 
\min_{e_{t+1}(\cdot),\dots,e_T(\cdot)}\mathsf{V}_t^\dagger(e_{t+1}(\cdot),\dots,e_T(\cdot))
\ \leq\ 
\min_{e_{s+1}(\cdot),\dots,e_T(\cdot)}\mathsf{V}_s^\dagger(e_{s+1}(\cdot),\dots,e_T(\cdot));
\end{align*}
and (ii) for any $\epsilon > 0$ and $d_\epsilon$ with $\min_{x\in\mathcal{X},a\in\mathcal{A}}\rho(d_{\epsilon}|x,a) \geq 1-\epsilon$, $\mathsf{V} \leq \mathsf{V}^* + C(\frac{d_{\epsilon}}{T} + \epsilon)$, where $C$ does not depend on $T$, $\epsilon$ or $d_\epsilon$.  
\qed
\end{thm}

\section{Additional simulation results}

We provide additional simulation evidence in this section on the finite-sample performance of the proposed fully forward-looking CARA design, focusing on the optimized variance under the power maximization objective across three extrapolation methods: neutral, optimistic, and conservative. The results are presented in Figures \ref{fig-1: variance for different extrapolation, X=sex, T=4} (for $T=4$) and \ref{fig-2: variance for different extrapolation, X=sex, T=6} (for $T=6$) where covariate stratification is based biological sex only, and in Figures \ref{fig-3: variance for different extrapolation, X=sex WHO, T=4} and \ref{fig-4: variance for different extrapolation, X=sex WHO, T=6} where we use both biological sex and WHO clinical stage. 

We highlight a few key observations. First, under the conservative extrapolation method, the optimized variance tends to be large at the beginning and gradually decreases as the experiment progresses. This pattern aligns with Theorem \ref{SA-thm-7: Optimized variance and bound on efficiency loss} and reflects the ``minimax'' nature of the conservative approach. Specifically, the conservative method initially assumes that none of the missing outcomes will become available, leading to high variance early on. However, as more data are collected, the sequentially learned delay mechanism helps reduce the variance objective function.

In contrast, under the optimistic extrapolation method, the variance bound exhibits the opposite trend. Because the optimistic method assumes that missing outcomes will become available after one period, the initial variance tends to be small. Yet as information about the actual delay mechanism accumulates, the experimenter must adjust their expectation, leading to an increase in the variance bound. Finally, the neutral method, which uses linear extrapolation, shows relatively stable variance bounds over time. Encouragingly, we find that the final performance of our procedure, measured by the variance at the end of the experiment, is relatively robust across extrapolation methods. For comparison, the optimized oracle variance is also shown in the figures as a horizontal dashed line.

\clearpage

\begin{figure}[!th]
    \centering
    \includegraphics[width=0.9\linewidth]{Revision_figures/supp-fig-1-variance-extrapolation-t4.pdf}
    \caption{Optimized variance for the power maximization objective. The three panels correspond to the conservative, the neutral, and the optimistic approach of extrapolating the estimated delay mechanism. Horizon: $T=4$. Covariate stratification is based on biological sex. 
    }
    \label{fig-1: variance for different extrapolation, X=sex, T=4}
\end{figure}

\begin{figure}[!th]
    \centering
    \includegraphics[width=0.9\linewidth]{Revision_figures/supp-fig-2-variance-extrapolation-t6.pdf}
    \caption{Optimized variance for the power maximization objective. The three panels correspond to the conservative, the neutral, and the optimistic approach of extrapolating the estimated delay mechanism. Horizon: $T=6$. Covariate stratification is based on biological sex. 
    }
    \label{fig-2: variance for different extrapolation, X=sex, T=6}
\end{figure}

\clearpage

\begin{figure}[!th]
    \centering
    \includegraphics[width=0.9\linewidth]{Revision_figures/supp-fig-3-variance-extrapolation-t4.pdf}
    \caption{Optimized variance for the power maximization objective. The three panels correspond to the conservative, the neutral, and the optimistic approach of extrapolating the estimated delay mechanism. Horizon: $T=4$. Covariate stratification is based on biological sex and WHO stage. 
    }
    \label{fig-3: variance for different extrapolation, X=sex WHO, T=4}
\end{figure}

\begin{figure}[!th]
    \centering
    \includegraphics[width=0.9\linewidth]{Revision_figures/supp-fig-4-variance-extrapolation-t6.pdf}
    \caption{Optimized variance for the power maximization objective. The three panels correspond to the conservative, the neutral, and the optimistic approach of extrapolating the estimated delay mechanism. Horizon: $T=6$. Covariate stratification is based on biological sex and WHO stage. 
    }
    \label{fig-4: variance for different extrapolation, X=sex WHO, T=6}
\end{figure}

\clearpage

\section{Proofs}

\subsection{Proof of Theorem \ref{SA-thm-1: Semiparametric efficiency bound}}

We omit the subscript $i$ in the following derivation to save notation. We break down our proof into three steps: (1) characterizing the tangent space, (2) expressing the causal parameter $\tau$ and computing the pathwise derivative, and (3) projecting the resulting influence function onto the tangent space to obtain the efficient influence function. See \cite{van2000asymptotic} and \cite{tsiatis2006semiparametric} for textbook treatments of semiparametric efficiency theory.

\noindent\textbf{Step 1.} We first find the tangent space. To this end, we denote the joint density of the observed data as $ f_{\eta}(X, A, D, Y, S)$, where the subscript $\eta $ is adopted to represent and parameterize a specific one dimensional parametric submodel passing through the truth: 
\begin{align*}
    & f_{\eta}(X, A, D, Y, S)\\
    = & p_{\eta}(X)\prod_{t=1}^T\Bigg\{ r_{t,\eta}\prod_{a\in\mathcal{A}}\Big[e_{t,\eta}(a|X)\Big(\rho_{\eta}(T-t|X,a)f_{Y(a)|X,\eta}(Y|X)\Big)^{\Indicator_{(D\leq T-t)}}\\
    &\qquad\qquad\qquad\qquad\qquad\qquad\Big(1-\rho_{\eta}(T-t|X,a)\Big)^{\Indicator_{(D> T-t)}}\Big]^{\Indicator_{(A=a)}} \Bigg\}^{\Indicator_{(S = t)}} .
\end{align*}
Here, we recall that $r_{t,\eta}$ is the enrollment frequency of stage $t$, $p_{\eta}(\cdot)$ is the unconditional distribution of the covariates, $e_{t,\eta}(\cdot|\cdot)$ is the treatment assignment probability, $\rho_\eta(\cdot|\cdot,\cdot)$ is the delay mechanism, and $f_{Y(a)|X,\eta}(\cdot|\cdot)$ is the conditional distribution of the potential outcomes. We next write the above joint density as
\begin{align*}
   &  f_{\eta}(X, A, D, Y, S)\\
   = & p_{\eta}(X)  \times \prod_{a\in\mathcal{A}}\Big(f_{Y(a)|X,\eta}(Y|X)\Big)^{\Indicator_{(A=a)}\sum_{t=1}^T \Indicator_{(S = t)}\Indicator_{(D\leq T-t)}} \\
   &\quad \times \Big(\prod_{t=1}^{T}r_{t,\eta}^{\Indicator_{(S = t)}}\Big)\\
    &\quad \times  \prod_{t=1}^T \prod_{a\in\mathcal{A}}\Big(e_{t,\eta}(a|X)\Big)^{\Indicator_{(S = t)}\Indicator_{(A=a)}}  \\
    &\quad \times \prod_{t=1}^T\prod_{a\in\mathcal{A}} \Big(\rho_{\eta}(T-t|X,a)\Big)^{\Indicator_{(S = t)}\Indicator_{(A=a)}\Indicator_{(D\leq T-t)}} \Big(1-\rho_{\eta}(T-t|X,a)\Big)^{\Indicator_{(S = t)}\Indicator_{(A=a)}\Indicator_{(D> T-t)}}.
\end{align*}
The tangent space consists of scores for different submodels. Since each component of the joint distribution can independently vary across submodels, the tangent space takes the form of a direct sum of the following tangent subspaces; each tangent subspace corresponds to scores computed for different components in the above joint density:
\begin{align*}
    \mathcal{T}_{X} &= \Big\{ \mathfrak{f}_X(X):\ \Expectation[\mathfrak{f}_X(X)]=0 \Big\};\\
    \mathcal{T}_{Y(a)} &= \Big\{\Big(\Indicator_{(A=a)}\sum_{t=1}^T \Indicator_{(S = t)}\Indicator_{(D\leq T-t)}\Big)\mathfrak{g}_{a}(Y|X):\ \Expectation[\mathfrak{g}_{a}(Y(a)|X)|X]=0\Big\}\quad \text{for }a\in\mathcal{A}=\{0,1\};\\
    \mathcal{T}_{S} &= \Big\{\sum_{t=1}^{T}\frac{\Indicator_{(S = t)}}{r_t} \mathfrak{a}_t:\ \sum_{t=1}^T\mathfrak{a}_t=0\Big\};\\
    \mathcal{T}_{A,t} &= \Big\{\Indicator_{(S = t)}\Big(\frac{\Indicator_{(A=1)}}{e_t(1|X)} - \frac{\Indicator_{(A=0)}}{e_t(0|X)}\Big)\mathfrak{e}_{t}(X):\ \mathfrak{e}_{t}(\cdot)\ \text{general function}\Big\};\\
    \mathcal{T}_{D,a,t} &= \Big\{\Indicator_{(S = t)}\Indicator_{(A = a)}\Big(\frac{\Indicator_{(D\leq T-t)}}{\rho(T-t|X,a)} - \frac{\Indicator_{(D> T-t)}}{1-\rho(T-t|X,a)}\Big)\mathfrak{p}_{a,T-t}(X):\ \mathfrak{p}_{a,T-t}(\cdot)\ \text{general function}\Big\}\\
    &\ \hskip0.75\textwidth\text{for }a\in\mathcal{A}=\{0,1\}.
\end{align*}
In the above, $\mathfrak{f}_X(\cdot)$, $\mathfrak{g}_{a}(\cdot|\cdot)$, $\mathfrak{a}_t$, $\mathfrak{e}_{t}(\cdot)$, and $\mathfrak{p}_{a,T-t}(\cdot)$ denote general functions satisfying certain restrictions specified in each of the tangent subspaces. For example, $\mathcal{T}_{X}$ consists of functions of $X$ that are mean zero.\footnote{Strictly speaking, elements of a tangent space should also have finite second moment. We dropped such qualifications in the definition of the subspaces to save notation and conserve space. See \cite{van2000asymptotic} for additional technical details.} 

It is routine to verify that the above spaces are mutually orthogonal (i.e., elements in the tangent subspaces are uncorrelated), which justifies that the whole tangent space can be written as their direct sum. 

\noindent\textbf{Step 2.} In this step, consider the estimand $\mu_1(a)=\Expectation[Y(a)]$ (for example, set $a=1$ to convey the main idea). We will take pathwise derivative of $\mu_1(a)$ and then find an influence function. We start from the following
\begin{align*}
    \mu_{1,\eta}(a) = \Expectation_\eta\Bigg[ \frac{\Indicator_{(S = 1)}\Indicator_{(A=a)}\Indicator_{(D\leq T-1)}Y}{r_{1,\eta}e_{1,\eta}(a|X)\rho_{\eta}(T-1|X,a)}\Bigg],
\end{align*}
and again, $\eta$ parameterizes a particular one-dimensional submodel. The pathwise derivative is 
\begin{align*}
\frac{\partial \mu_{1,\eta}(a) }{\diff \eta}\Bigg|_{\eta=0} &= \Expectation\Bigg[ \left(\frac{\Indicator_{(S = 1)}\Indicator_{(A=a)}\Indicator_{(D\leq T-1)}Y}{r_1e_{1}(a|X)\rho(T-1|X,a)} - \mu_1(a)\right)\mathfrak{s}(X,A,D,Y,S)\Bigg]\\
\tag{I}&- \Expectation\Bigg[ \frac{\Indicator_{(S = 1)}\Indicator_{(A=a)}\Indicator_{(D\leq T-1)}Y}{\Big(r_1e_{1}(a|X)\rho(T-1|X,a)\Big)^2}e_{1}(a|X)\rho(T-1|X,a)\mathfrak{a}_1\Bigg]\\
\tag{II}&- \Expectation\Bigg[ \frac{\Indicator_{(S = 1)}\Indicator_{(A=a)}\Indicator_{(D\leq T-1)}Y}{\Big(r_1e_{1}(a|X)\rho(T-1|X,a)\Big)^2}r_1\rho(T-1|X,a)\mathfrak{e}_{1}(X)\Bigg]\\
\tag{III}&- \Expectation\Bigg[ \frac{\Indicator_{(S = 1)}\Indicator_{(A=a)}\Indicator_{(D\leq T-1)}Y}{\Big(r_1e_1(a|X)\rho(T-1|X,a)\Big)^2}r_1e_{1}(a|X)\mathfrak{p}_{a,T-1}(X)\Bigg],
\end{align*}
and $\mathfrak{s}(\cdot,\cdot,\cdot,\cdot,\cdot)$ is the score for this particular parametric submodel. Then
\begin{align*}
\text{(I)}
&= \Expectation\Bigg[ -\frac{\Indicator_{(S = 1)}\mu_1(X,a)}{r_1^2}\mathfrak{a}\Bigg] = - \Expectation\Bigg[ \frac{\Indicator_{(S = 1)}\mu_1(X,a)}{r_1}\left(\sum_{t=1}^{T}\frac{\Indicator_{(S = t)}}{r_t} \mathfrak{a}_t\right)\Bigg]\\
&=  \Expectation\Bigg[ -\frac{(\Indicator_{(S = 1)}-r_1)\mu_1(X,a)}{r_1}\left(\sum_{t=1}^{T}\frac{\Indicator_{(S = t)}}{r_t} \mathfrak{a}_t\right)\Bigg]
= \Expectation\Bigg[ -\frac{(\Indicator_{(S= 1)}-r_1)\mu_1(X,a)}{r_1}\mathfrak{s}(X,A,D,Y,S)\Bigg].
\end{align*}
Similarly
\begin{align*}
\text{(II)}
&= \Expectation\Bigg[ -\frac{\Indicator_{(S = 1)}\Indicator_{(A=a)} \mu_1(X,a)}{r_1e_{1}(a|X)^2}\mathfrak{e}_{1}(X)\Bigg]\\
&= \Expectation\Bigg[ -\frac{\Indicator_{(S = 1)}\Indicator_{(A=a)}  \mu_1(X,a)}{r_1 e_{1}(a|X)}\Indicator_{(S = 1)}\left(\frac{\Indicator_{(A=1)}}{e_1(1|X)} - \frac{\Indicator_{(A=0)}}{e_1(0|X)}\right)\mathfrak{e}_{1}(X)\Bigg]\\
&= \Expectation\Bigg[ -\frac{\Indicator_{(S = 1)}(\Indicator_{(A=a)}-e_{1}(a|X)) \mu_1(X,a)}{r_1e_{1}(a|X)}\Indicator_{(S = 1)}\left(\frac{\Indicator_{(A=1)}}{e_1(1|X)} - \frac{\Indicator_{(A=0)}}{e_1(0|X)}\right)\mathfrak{e}_{1}(X)\Bigg]\\
&= \Expectation\Bigg[ -\frac{\Indicator_{(S = 1)}(\Indicator_{(A=a)}-e_{1}(a|X)) \mu_1(X,a)}{r_1e_{1}(a|X)}\mathfrak{s}(X,A,D,Y,S)\Bigg].
\end{align*}
Finally, 
\begin{align*}
\text{(III)}
&= \Expectation\Bigg[ -\frac{\Indicator_{(S = 1)}\Indicator_{(A=a)}(\Indicator_{(D\leq T-1)} - \rho(T-1|X,a)) \mu_1(X,a)}{r_1e_{1}(a|X)\rho(T-1|X,a)}\mathfrak{s}(X,A,D,Y,S)\Bigg].
\end{align*}
Therefore by collecting terms, 
\begin{align*}
\frac{\partial \mu_{1,\eta}(a) }{\diff \eta}\Bigg|_{\eta=0} &= \Expectation\left[\Big(\phi_1(X,A,D,Y,S)+\phi_2(X)\Big)\mathfrak{s}(X,A,D,Y,S)\right],
\end{align*}
where
\begin{align*}
    \phi_1(X,A,D,Y,S) &= \frac{\Indicator_{(S = 1)}\Indicator_{(A=a)}\Indicator_{(D\leq T-t)}(Y-\mu_1(X,a))}{r_1e_{1}(a|X)\rho(T-t|X,a)},\quad
    \phi_2(X) = \mu_1(X,a) - \mu_1(a).
\end{align*}
The above shows that the pathwise derivative can be written as an inner product with the score. As a result, $\phi_1(X,A,D,Y,S) + \phi_2(X)$ gives an influence function. The problem, however, is that $\phi_1$ is not in the tangent space (note that $\phi_2$ is in $\mathcal{T}_X$). 

\noindent\textbf{Step 3}. In this step, we project $\phi_1$ onto the tangent space to obtain the efficient influence function. To complete the efficient influence function, we need to find the projection of the first term to $\mathcal{T}_{Y(a)}$. (Note that the other tangent spaces are irrelevant, as $\phi_1$ is already orthogonal to them.) The projection is uniquely determined by
\begin{align*}
    &\Expectation\Bigg[\left(\phi_1(X,A,D,Y,S) - \Big(\Indicator_{(A=a)}\sum_{t=1}^T \Indicator_{(S = t)}\Indicator_{(D\leq T-t)}\Big)\mathfrak{g}_{a}(Y|X)\right)
     \cdot\Big(\Indicator_{(A=a)}\sum_{t=1}^T \Indicator_{(S = t)}\Indicator_{(D\leq T-t)}\Big)\mathfrak{g}_{a}(Y|X)\Bigg]=0.
\end{align*}
First, we have
\begin{align*}
&\ \Expectation\Bigg[\phi_1(X,A,D,Y,S) \Big(\Indicator_{(A=a)}\sum_{t=1}^T \Indicator_{(S = t)}\Indicator_{(D\leq T-t)}\Big)\mathfrak{g}_{a}(Y|X)\Bigg]
= \Expectation[(Y(a)-\mu_1(X,a))\mathfrak{g}_{a}(Y|X)].
\end{align*}
We also have
\begin{align*}
&\ \Expectation\left[\left( \Big(\Indicator_{(A=a)}\sum_{t=1}^T \Indicator_{(S = t)}\Indicator_{(D\leq T-t)}\Big)\mathfrak{g}_{a}(Y|X)\right)^2\right]
= \Expectation\left[\left(\sum_{t=1}^T r_te_t(a|X)\rho(T-t|X,a)\right)\Expectation[\mathfrak{g}_{a}(Y(a)|X)^2|X]\right].
\end{align*}
Then we can simply set
\begin{align*}
\mathfrak{g}_{a}(Y(a)|X) &= \frac{Y(a)-\mu_1(X,a)}{\sum_{t=1}^T r_te_t(a|X)\rho(T-t|X,a)}.
\end{align*}

\subsection{Proof of Lemma \ref{SA-lem-1: Consistent estimation of mean potential outcomes}}

We start by writing
\begin{align*}
    \hat{\mu}_{s,T}(x,a) - \mu_s(x,a) &= \frac{ \sum_{i=1}^{N} \Indicator_{(X_{i}=x,A_{i}=a,D_{i}+S_i\leq T)}\Big(Y_{i}^s - \mu_s(x,a)\Big)}{ \sum_{i=1}^{N} \Indicator_{(X_{i}=x,A_{i}=a,D_{i}+S_i\leq T)}\phantom{\Big(Y_{i}^s - \mu_s(x,a)\Big)}}\\
    &= \Bigg(\frac{1}{N}\sum_{i=1}^{N} \Indicator_{(X_{i}=x,A_{i}=a,D_{i}+S_i\leq T)}\Bigg)^{-1}\Bigg(\frac{1}{N}\sum_{i=1}^{N} \Indicator_{(X_{i}=x,A_{i}=a,D_{i}+S_i\leq T)}\Big(Y_{i}^s - \mu_s(x,a)\Big)\Bigg).
\end{align*}
Next, recall that individuals are ordered by the stage in which they are enrolled. Letting $N_t = \sum_{s=1}^t n_s$, we have
\begin{align*}
\frac{1}{N}\sum_{i=1}^{N} \Indicator_{(X_{i}=x,A_{i}=a,D_{i}+S_i\leq T)} &= \frac{1}{N}\sum_{t=1}^T\sum_{i=N_{t-1}+1}^{N_t} \Indicator_{(X_{i}=x,A_{i}=a,D_{i}\leq T-t)}\\
&= \frac{1}{N}\sum_{t=1}^T\sum_{i=N_{t-1}+1}^{N_t} \Big(\Indicator_{(X_{i}=x,A_{i}=a,D_{i}\leq T-t)} - p(x)\hat{e}^*_t(a|x)\rho(T-t|x,a)\Big)\\
&\qquad\qquad+ \frac{1}{N}\sum_{t=1}^T\sum_{i=N_{t-1}+1}^{N_t} p(x)\hat{e}^*_t(a|x)\rho(T-t|x,a).
\end{align*}
Given our assumption that $p(x)$, $\rho(0|x,a)$, and the treatment probabilities are bounded from below, 
\begin{align*}
\frac{1}{N}\sum_{t=1}^T\sum_{i=N_{t-1}+1}^{N_t} p(x)\hat{e}^*_t(a|x)\rho(T-t|x,a) \succsim 1. 
\end{align*}
Next, we notice the variance satisfies
\begin{align*}
    &\Var\left[\frac{1}{N}\sum_{t=1}^T\sum_{i=N_{t-1}+1}^{N_t} \Big(\Indicator_{(X_{i}=x,A_{i}=a,D_{i}\leq T-t)} - p(x)\hat{e}^*_t(a|x)\rho(T-t|x,a)\Big)\right]\\
    =& \frac{1}{N^2}\sum_{t=1}^T\Expectation\left[\Var\left[\sum_{i=N_{t-1}+1}^{N_t} \Big(\Indicator_{(X_{i}=x,A_{i}=a,D_{i}\leq T-t)} - p(x)\hat{e}^*_t(a|x)\rho(T-t|x,a)\Big)\ \Bigg|\ \mathcal{H}_{t-1} \right]\right] = O\left(\frac{1}{N}\right). 
\end{align*}
As a result, 
\begin{align*}
\Bigg(\frac{1}{N}\sum_{i=1}^{N} \Indicator_{(X_{i}=x,A_{i}=a,D_{i}+S_i\leq T)}\Bigg)^{-1} = \Op(1). 
\end{align*}

To conclude the proof, we again compute the variance
\begin{align*}
    &\Var\left[\frac{1}{N}\sum_{i=1}^{N} \Indicator_{(X_{i}=x,A_{i}=a,D_{i}+S_i\leq T)}\Big(Y_{i}^s - \mu_s(x,a)\Big)\right]\\
    =&\Var\left[\frac{1}{N}\sum_{t=1}^T\sum_{i=N_{t-1}+1}^{N_t} \Indicator_{(X_{i}=x,A_{i}=a,D_{i}\leq T-t)}\Big(Y_{i}^s - \mu_s(x,a)\Big)\right]\\
    =& \frac{1}{N^2}\sum_{t=1}^T\Expectation\left[\Var\left[\sum_{i=N_{t-1}+1}^{N_t} \Indicator_{(X_{i}=x,A_{i}=a,D_{i}\leq T-t)}\Big(Y_{i}^s - \mu_s(x,a)\Big)\ \Bigg|\ \mathcal{H}_{t-1} \right]\right] = O\left(\frac{1}{N}\right). 
\end{align*}
The bound following the last inequality is due to our assumption on the finite fourth moment of the potential outcomes. 

\subsection{Proof of Lemma \ref{SA-lem-2: Asymptotic normality of estimated mean potential outcomes}}

Recall that individuals are ordered by the stage in which they are enrolled, and therefore we let $N_t = \sum_{s=1}^t n_s$. We start with the denominator, which can be rewritten as
\begin{align*}
\frac{1}{N}\sum_{i=1}^{N} \Indicator_{(X_{i}=x,A_{i}=a,D_{i}+S_i\leq T)}    =&\frac{1}{N}\sum_{t=1}^T \sum_{i=N_{t-1}+1}^{N_t} \Indicator_{(X_{i}=x,A_{i}=a,D_{i}\leq T-t)}\\ 
    =& \frac{1}{N}\sum_{t=1}^T \sum_{i=N_{t-1}+1}^{N_t} \big(\Indicator_{(X_{i}=x,A_{i}=a,D_{i}\leq T-t)} - p(x)\hat{e}^*_t(a|x)\rho(T-t|x,a) \big)\\
    &\quad + p(x)\sum_{t=1}^T \frac{n_t}{N} \hat{e}^*_t(a|x)\rho(T-t|x,a). 
\end{align*}
Clearly the second term satisfies
\begin{align*}
p(x)\sum_{t=1}^T \frac{n_t}{N} \hat{e}^*_t(a|x)\rho(T-t|x,a) = p(x)\sum_{t=1}^T r_t e_t(a|x) \rho(T-t|x,a) + \op(1)
\end{align*}
by our condition on the consistency of treatment allocation. 

The first term is easily shown to be mean zero by a standard martingale calculation. Its variance is
\begin{align*}
&\Var\left[ \frac{1}{N}\sum_{t=1}^T \sum_{i=N_{t-1}+1}^{N_t} \big(\Indicator_{(X_{i}=x,A_{i}=a,D_{i}\leq T-t)} - p(x)\hat{e}^*_t(a|x)\rho(T-t|x,a) \big) \right]\\
=& \frac{1}{N^2}\sum_{t=1}^T n_t \Expectation\Big[p(x)\hat{e}^*_t(a|x)\rho(T-t|x,a)\Big(1-p(x)\hat{e}^*_t(a|x)\rho(T-t|x,a)\Big)\Big] = O\left(\frac{1}{N}\right).  
\end{align*}
As a result, we showed that
\begin{align*}
\frac{1}{N}\sum_{t=1}^T \sum_{i=N_{t-1}+1}^{N_t} \Indicator_{(X_{i}=x,A_{i}=a,D_{i}\leq T-t)} &= p(x)\sum_{t=1}^T r_t e_t(a|x) \rho(T-t|x,a) + \op(1).
\end{align*}

We now analyze the numerator. It can be rewritten as
\begin{align*}
\frac{1}{\sqrt{N}}\sum_{t=1}^T \sum_{i=N_{t-1}+1}^{N_t}\Indicator_{(X_{i}=x,A_{i}=a,D_{i}\leq T-t)}\big(Y_{i} - \mu_1(x,a)\big).
\end{align*}
To establish asymptotic normality, we apply the martingale central limit theorem, which requires computing the conditional variance and verify a Lindeberg-Feller condition \citep{hall2014martingale}. The conditional variance is given by
\begin{align*}
\frac{1}{\sqrt{N}}\sum_{t=1}^T \sum_{i=N_{t-1}+1}^{N_t} \Var\Big[ \Indicator_{(X_{i}=x,A_{i}=a,D_{i}\leq T-t)}\big(Y_{i} - \mu_1(x,a)\big) \Big| \{(X_{j},A_{j},D_{j},Y_{j}):1\leq j < i\}\Big].
\end{align*}
Next, we notice that individuals enrolled in the same stage are independent of each other, and the treatment assignment only depends on accrued information up to the previous stage. As a result, the above reduces to
\begin{align*}
&\frac{1}{\sqrt{N}}\sum_{t=1}^T \sum_{i=N_{t-1}+1}^{N_t} \Var\Big[ \Indicator_{(X_{i}=x,A_{i}=a,D_{i}\leq T-t)}\big(Y_{i} - \mu_1(x,a)\big) \Big| \{(X_{j},A_{j},D_{j},Y_{j}):1\leq j \leq N_{t-1}\}\Big]\\
=&\frac{1}{N}\sum_{t=1}^T n_t \Var\Big[ \Indicator_{(X_{i}=x,A_{i}=a,D_{i}\leq T-t)}\big(Y_{i} - \mu_1(x,a)\big) \Big| \mathcal{H}_{t-1}\Big]\\
=& \sum_{t=1}^T\frac{n_t}{N} p(x) \hat{e}^*_t(a|x) \rho(T-t|x,a) \sigma^2(x,a) 
= p(x)\sigma^2(x,a)\sum_{t=1}^T r_t  e_t(a|x) \rho(T-t|x,a)  + \op(1).
\end{align*}
Finally, we verify the Lindeberg-Feller condition for the numerator with fourth conditional moments:
\begin{align*}
&\frac{1}{N^2}\sum_{t=1}^T \sum_{i=N_{t-1}+1}^{N_t} \Expectation\left[\Big|\Indicator_{(X_{i}=x,A_{i}=a,D_{i}\leq T-t)}\big(Y_{i} - \mu_1(x,a)\big)\Big|^4\ \Big|\ \{(X_{j},A_{j},D_{j},Y_{j}):1\leq j < i\}\right]\\
=&\frac{1}{N^2}\sum_{t=1}^T \sum_{i=N_{t-1}+1}^{N_t} \Expectation\left[\Big|\Indicator_{(X_{i}=x,A_{i}=a,D_{i}\leq T-t)}\big(Y_{i} - \mu_1(x,a)\big)\Big|^4\ \Big|\ \mathcal{H}_{t-1}\right] \precsim  \frac{1}{N^2}\sum_{t=1}^T n_t  = o(1).
\end{align*}

\subsection{Proof of Theorem \ref{SA-thm-2: Asymptotic normality of estimated subgroup treatment effects}}

In the proof of Lemma \ref{SA-lem-2: Asymptotic normality of estimated mean potential outcomes}, we notice that the term for the control units, 
\[
\Indicator_{(X_{i}=x,A_{i}=0,D_{i}\leq T-t)}\big(Y_{i} - \mu_1(x,0)\big),
\] 
and that for the treated units, 
\[
\Indicator_{(X_{i}=x,A_{i}=1,D_{i}\leq T-t)}\big(Y_{i} - \mu_1(x,1)\big),\] 
have zero conditional covariance due to the indicators $\Indicator_{(A_{i}=0)}$ and $\Indicator_{(A_{i}=1)}$. As a result, $\hat{\mu}_{1,T}(x,1)$ and $\hat{\mu}_{1,T}(x,0)$ are jointly asymptotically normal with a zero covariance, meaning that they are asymptotically independent:
\begin{align*}
\begin{bmatrix}
\sqrt{N}\big( \hat{\mu}_{1,T}(x,0) - \mu_1(x,0) \big)\\
\sqrt{N}\big( \hat{\mu}_{1,T}(x,1) - \mu_1(x,1) \big)
\end{bmatrix}
\toDist \mathcal{N}\left(0,  \begin{bmatrix}
\frac{\sigma^2(x,0)}{p(x)\sum_{t=1}^T r_t e_t(0|x) \rho(T-t|x,0)} & 0\\
0 & \frac{\sigma^2(x,1)}{p(x)\sum_{t=1}^T r_t e_t(1|x) \rho(T-t|x,1)}
\end{bmatrix} \right).
\end{align*}
The conclusion of the theorem then follows from recognizing $\hat{\tau}_T(x) = \hat{\mu}_{1,T}(x,1) - \hat{\mu}_{1,T}(x,0)$, whose asymptotic distribution is normal with a variance being the sum of the two variance terms in the previous display.

\subsection{Proof of Theorem \ref{SA-thm-3: Asymptotic normality of estimated treatment effects}}

We start by writing the centered and scaled estimator as
\begin{align*}
&\sqrt{N}\big(\hat{\tau}_T - \tau\big) 
= \sum_{x\in \mathcal{X} } \hat{p}_T(x)\sqrt{N}\big(\hat{\tau}_T(x) - \tau(x)\big) + \sum_{x\in \mathcal{X} } \sqrt{N}\hat{p}_T(x)\big(\tau(x) - \tau\big).
\end{align*}
Next, we recall that $\sum_{x\in \mathcal{X}} p(x) \tau(x)  = \tau$, which implies
\begin{align*}
&\sqrt{N}\big(\hat{\tau}_T - \tau\big) 
= \underbrace{\sum_{x\in \mathcal{X} } \hat{p}_T(x)\sqrt{N}\big(\hat{\tau}_T(x) - \tau(x)\big)}_{\textstyle \text{(I)}} + \underbrace{\sum_{x\in \mathcal{X} } \sqrt{N}\big(\hat{p}_T(x) - p(x)\big)\big(\tau(x) - \tau\big)}_{\textstyle \text{(II)}},
\end{align*}
where we labeled the two terms above by (I) and (II), respectively. By Theorem \ref{SA-lem-2: Asymptotic normality of estimated mean potential outcomes} and its proof, term (I) can be written as
\begin{align*}
\text{(I)} &= \frac{1}{\sqrt{N}}\sum_{t=1}^T \sum_{i=N_{t-1}+1}^{N_t} \sum_{x\in\mathcal{X}}\frac{\Indicator_{(X_{i}=x,A_{i}=1,D_{i}\leq T-t)}}{\sum_{t=1}^T r_t e_t(1|x) \rho(T-t|x,1)}\big(Y_{i} - \mu_1(x,1)\big)\\
&\quad- \frac{1}{\sqrt{N}}\sum_{t=1}^T \sum_{i=N_{t-1}+1}^{N_t}  \sum_{x\in\mathcal{X}}\frac{\Indicator_{(X_{i}=x,A_{i}=0,D_{i}\leq T-t)}}{\sum_{t=1}^T r_t e_t(0|x) \rho(T-t|x,0)}\big(Y_{i} - \mu_1(x,0)\big) + \op(1).
\end{align*}
Term (II) expands into
\begin{align*}
\text{(II)} &= \frac{1}{\sqrt{N}}\sum_{t=1}^T \sum_{i=N_{t-1}+1}^{N_t}  \sum_{x\in\mathcal{X}} \big(\tau(x) - \tau\big)   \Indicator_{(X_{i}=x)} .
\end{align*}
We notice that it is already mean zero. To summarize, we have
\begin{align*}
&\sqrt{N}\big(\hat{\tau}_T - \tau\big)  
= \frac{1}{\sqrt{N}}\sum_{t=1}^T \sum_{i=N_{t-1}+1}^{N_t}  \Big(u_{i} - v_{i} + w_{i}\Big) + \op(1),
\end{align*}
where 
\begin{align*}
u_{i} &= \sum_{x\in\mathcal{X}}\frac{\Indicator_{(X_{i}=x,A_{i}=1,D_{i}\leq T-t)}}{\sum_{t=1}^T r_t e_t(1|x) \rho(T-t|x,1)}\big(Y_{i} - \mu_1(x,1)\big)\\
v_{i} &=\sum_{x\in\mathcal{X}}\frac{\Indicator_{(X_{i}=x,A_{i}=0,D_{i}\leq T-t)}}{\sum_{t=1}^T r_t e_t(0|x) \rho(T-t|x,0)}\big(Y_{i} - \mu_1(x,0)\big)\\
w_{i} &= \sum_{x\in\mathcal{X}} \big(\tau(x) - \tau\big)   \Indicator_{(X_{i}=x)}.
\end{align*}

We will compute the conditional variance for each term. For the term involving $u_i$, one has 
\begin{align*}
&\frac{1}{N}\sum_{t=1}^T \sum_{i=N_{t-1}+1}^{N_t} \Var\Big[u_i\Big| \{(X_{j},A_{j},D_{j},Y_{j}):1\leq j < i\}\Big] = \frac{1}{N}\sum_{t=1}^T \sum_{i=N_{t-1}+1}^{N_t} \Var\Big[u_i\Big| \mathcal{H}_{t-1}\Big] \\
=&\frac{1}{N}\sum_{t=1}^T \sum_{i=N_{t-1}+1}^{N_t} \sum_{x\in\mathcal{X}}\Var\Big[\frac{\Indicator_{(X_{i}=x,A_{i}=1,D_{i}\leq T-t)}}{\sum_{t=1}^T r_t e_t(1|x) \rho(T-t|x,1)}\big(Y_{i} - \mu_1(x,1)\big)\Big| \mathcal{H}_{t-1}\Big] \\
=& \sum_{t=1}^T \sum_{x\in\mathcal{X}} \frac{n_t}{N} \frac{p(x)\hat{e}^*_t(1|x)\rho(T-t|x,1)}{(\sum_{t=1}^T r_t e_t(1|x) \rho(T-t|x,1))^2}\sigma^2(x,1)
= \sum_{x\in\mathcal{X}} p(x) \frac{\sigma^2(x,1)}{\sum_{t=1}^T r_t e_t(1|x) \rho(T-t|x,1)} +\op(1),
\end{align*}
and for the term with $v_i$, one has
\begin{align*}
&\frac{1}{N}\sum_{t=1}^T \sum_{i=N_{t-1}+1}^{N_t}\Var\Big[v_i\Big| \{(X_{j},A_{j},D_{j},Y_{j}):1\leq j < i\}\Big] = \frac{1}{N}\sum_{t=1}^T \sum_{i=N_{t-1}+1}^{N_t}\Var\Big[v_i\Big| \mathcal{H}_{t-1}\Big] \\
=&\frac{1}{N}\sum_{t=1}^T \sum_{i=N_{t-1}+1}^{N_t}\sum_{x\in\mathcal{X}}\Var\Big[\frac{\Indicator_{(X_{i}=x,A_{i}=0,D_{i}\leq T-t)}}{\sum_{t=1}^T r_t e_t(0|x) \rho(T-t|x,0)}\big(Y_{i} - \mu_1(x,0)\big)\Big| \mathcal{H}_{t-1}\Big] \\
=& \sum_{t=1}^T \sum_{x\in\mathcal{X}} \frac{n_t}{N} \frac{p(x)\hat{e}^*_t(0|x)\rho(T-t|x,0)}{(\sum_{t=1}^T r_t e_t(0|x) \rho(T-t|x,0))^2}\sigma^2(x,0)\\
=& \sum_{x\in\mathcal{X}} p(x) \frac{\sigma^2(x,0)}{\sum_{t=1}^T r_t e_t(0|x) \rho(T-t|x,0)} +\op(1),
\end{align*}
and finally,
\begin{align*}
&\frac{1}{N}\sum_{t=1}^T \sum_{i=N_{t-1}+1}^{N_t}  \Var\Big[w_i \Big|\{(X_{j},A_{j},D_{j},Y_{j}):1\leq j < i\}\Big] = \frac{1}{N}\sum_{t=1}^T \sum_{i=N_{t-1}+1}^{N_t}  \Var\Big[w_i \Big|\mathcal{H}_{t-1}\Big] \\
=&\frac{1}{N}\sum_{t=1}^T \sum_{i=N_{t-1}+1}^{N_t} \sum_{x\in\mathcal{X}} \Var\Big[\big(\tau(x) - \tau\big) \Indicator_{(X_{i}=x)} \Big|\mathcal{H}_{t-1}\Big] \\
=& \frac{1}{N}\sum_{t=1}^T \sum_{i=N_{t-1}+1}^{N_t} \sum_{x\in\mathcal{X}} \big(\tau(x) - \tau\big)^2 p(x) = \sum_{x\in\mathcal{X}} \big(\tau(x) - \tau\big)^2 p(x).
\end{align*}
Next, we notice that $u_i$, $v_i$, and $w_i$ are (conditionally) uncorrelated. Specifically, 
\begin{align*}
\Expectation[u_iv_i|\{(X_{j},A_{j},D_{j},Y_{j}):1\leq j < i\}] = 0
\end{align*}
due to the presence of the indicator $\Indicator_{(A_i=1)}$ in $u_i$ and the indicator $\Indicator_{(A_i=0)}$ in $v_i$. The zero correlation between $u_i$ and $w_i$ follows from
\begin{align*}
&\Expectation[u_iw_i|\{(X_{j},A_{j},D_{j},Y_{j}):1\leq j < i\}] = \Expectation[u_iw_i|\mathcal{H}_{t-1}]\\
=& \sum_{x\in\mathcal{X}} \Expectation\Big[\frac{\Indicator_{(X_{i}=x,A_{i}=1,D_{i}\leq T-t)}}{\sum_{t=1}^T r_t e_t(1|x) \rho(T-t|x,1)}\big(Y_{i} - \mu_1(x,1)\big)\big(\tau(x) - \tau\big)  \Big| \mathcal{H}_{t-1} \Big] = 0
\end{align*}
by the law of iterated expectation, since $Y_{i} - \mu_1(x,1)$ is mean zero by conditioning on $\mathcal{H}_{t-1}$ and $\Indicator_{(X_{i}=x,A_{i}=1,D_{i}\leq T-t)}$. Similarly, $\Expectation[v_iw_i|\{(X_{j},A_{j},D_{j},Y_{j}):1\leq j < i\}] = 0$.

To complete the proof of asymptotic normality, one needs to verify a conditional Lindeberg-Feller condition. See, for example, the proof of Lemma \ref{SA-lem-2: Asymptotic normality of estimated mean potential outcomes} for an illustration.

\subsection{Proof of Theorem \ref{SA-thm-4: Statistical inference}}

By the standard law of large numbers, we have
\begin{align*}
    \hat{p}_T(x) = p(x) + \Op\left(\sqrt{\frac{1}{N}}\right). 
\end{align*}
Then from the proof of Lemma \ref{SA-lem-2: Asymptotic normality of estimated mean potential outcomes}, 
\begin{align*}
\hat{\mathsf{e}}_T(x,a) &= \sum_{t=1}^T r_t e_t(a|x) \rho(T-t|x,a) + \Op\left(\sqrt{\frac{1}{N}}\right). 
\end{align*}
Finally, consistency of the estimated conditional variance, $\hat{\sigma}_T^2(x,a)$, directly follows from Lemma \ref{SA-lem-1: Consistent estimation of mean potential outcomes}. This establishes the consistency of the variance estimator. 

\subsection{\textsf{R} code for Algorithm \ref{algo:dim reduction}}\label{section: R code}

\begin{lstlisting}
#############################################################################
# The program minimizes the following objective function
# 
#                     c1               
#     ----------------------------------
#      a0 + a1*e1 + a2*e2 + ... + aJ*eJ 
# 
#                     c0
#  +  ---------------------------------------------- 
#      b0 + b1*(1-e1) + b2*(1-e2) + ... + bJ*(1-eJ) 
#
# subject to the constraints
#
#   low <= e1,e_2,...,eJ <= up 
#############################################################################
# c1 and c0 correspond to conditional variances 
# e1 represents the treatment allocation for the next period to be optimized
# a1,...,aJ and b1,...,bJ consist of the estimated delay mechanism and the 
#   stage-specific enrollment frequencies
# a0 and b0 are introduced to allow either a pilot, or to capture allocations 
#   in previous stages
#############################################################################
# Parameters
#   c1 and c0, as described above
#   a0 and b0: as described above
#   a and b, vectors of a1,...,aJ and b1,...,bJ
#   low and up: as described above
# Returned values:
#   allocation: optimized allocations
#   objective: minimized objective function
#############################################################################

delayedAssign <- function(c1, c0, a0 = 0, b0 = 0, a, b, low, up) {
  
  # number of stages
  Nstage <- length(a) 
  # sort stages according to ratio a/b
  stageIndex <- sort(a/b, index.return = TRUE)$ix
  a <- a[stageIndex]; b <- b[stageIndex]
  
  # initialization
  allocation <- matrix(NA, nrow = Nstage, ncol = Nstage)
  for (i in 1:Nstage) for (j in 1:Nstage) {
    allocation[i, j] <- up * (j > i) + low * (j < i)
  }
  objective <- rep(0, Nstage)
  
  # optimization
  for (i in 1:Nstage) {
    if (i == 1 & i == Nstage) { # only one stage to optimize
      objFunc <- function(x) {
        (c1 / (a0 + a[i] * x)) + (c0 / (b0 + b[i] * (1-x)))
      }
    } else {
      objFunc <- function(x) {
        (c1 / (a0 + sum(allocation[i, -1*i] * a[-1*i]) + a[i] * x)) + 
          (c0 / (b0 + sum((1-allocation[i, -1*i]) * b[-1*i]) + b[i] * (1-x)))
      }
    }
    
    temp <- optimize(objFunc, lower = low, upper = up, maximum = FALSE)
    objective[i] <- temp$objective; allocation[i, i] <- temp$minimum
  }
  allocation[, stageIndex] <- allocation
  
  # compute optimized allocation
  allocation <- allocation[which.min(objective), ]
  objective <- min(objective)
  return(list(allocation = allocation, objective = objective)) 
}
\end{lstlisting}

\subsection{Proof of Lemma \ref{SA-lem-3: unique solution beta prime beta in the reduced problem}}

The objective function $\mathsf{V}(\alpha,\beta)$ is strictly convex, and the domain of the optimization problem $\mathcal{V}$ is convex. Therefore, the solution is unique. We also notice that $\mathsf{V}$ is strictly decreasing with respect to the partial ordering of $\mathbb{R}^2$, which means the minimizer must lie on the upper boundary of $\mathcal{V}$.

\subsection{Proof of Theorem \ref{SA-thm-5: solution to the original optimization problem}}

The first claim of the theorem follows from the construction of $\mathcal{V}$. To show the second claim, we notice that the line segments connecting $(\alpha_j,\beta_j)$ all have different slopes. Therefore, the solution of the minimization problem either will  be a vertex or will lie on a line segment, and in the latter scenario, the line segment is unique. 

\subsection{Proof of Corollary \ref{SA-coro:Uniqueness of optimized allocations}}

This result follows directly from Lemma \ref{SA-lem-3: unique solution beta prime beta in the reduced problem} and Theorem \ref{SA-thm-5: solution to the original optimization problem}. 

\subsection{Proof of Theorem \ref{SA-thm-6: Consistent treatment allocation}}

We first note that minimizing the objective function $\hat{\mathsf{V}}_t^{\mathtt{forward}}$ is equivalent to minimizing 
\begin{align*}
\hat{\mathsf{V}}_t^{\mathtt{forward}}(e_{t+1},\dots,e_{T}|x) &= \frac{ \hat{\sigma}_{t}^2(x,1) }
 { \sum\limits_{\ell=1}^{t}\hat{\mathsf{b}}_\ell'(x)\hat{e}_\ell^*(1|x)+ \sum\limits_{\ell=t+1}^{T}\hat{\mathsf{b}}_{\ell}'(x) e_{\ell}} +\frac{ \hat{\sigma}_{t}^2(x,0) }
 {\sum\limits_{\ell=1}^{t} \hat{\mathsf{b}}_{\ell}(x) \hat{e}_\ell^*(0|x)+ \sum\limits_{\ell=t+1}^{T}\hat{\mathsf{b}}_{\ell}(x) (1-e_{\ell})}.
\end{align*}
That is, the optimization can be done separately for each subgroup $x\in\mathcal{X}$. Similarly, we define
\begin{align*}
    \mathsf{V}_t^\dagger(e_{t+1},\dots,e_T|x) &= \frac{\sigma^2(x,1)}{\sum\limits_{\ell=1}^{t}r_\ell \rho_t^\dagger(T-\ell|x,1)e_\ell^\dagger(1|x) +  \sum\limits_{\ell=t+1}^{T}r_\ell \rho_t^\dagger(T-\ell|x,1)e_\ell} \\
    & +  \frac{\sigma^2(x,0)}{\sum\limits_{\ell=1}^{t}r_\ell \rho_t^\dagger(T-\ell|x,0)e_\ell^\dagger(0|x) +  \sum\limits_{\ell=t+1}^{T}r_\ell \rho_t^\dagger(T-\ell|x,0)(1-e_\ell)},
\end{align*}
which is the large-sample analogue of $\hat{\mathsf{V}}_t^{\mathtt{forward}}(e_{t+1},\dots,e_{T}|x)$. In the remaining of this proof, we will fix some $x\in\mathcal{X}$ and show the consistency of $\hat{e}^*_{t+1}(1|x)$ for $t=1,2,\dots,T$. 

To start, the conclusion holds for $t=1$ by design: $\hat{e}_1^*(1|x) = {e}_1^\dagger(1|x) = 1/2$. Now assume the conclusion holds for $\ell = 1,2,\dots,t$. Then by employing the same technique used in the proof of Lemma \ref{SA-lem-1: Consistent estimation of mean potential outcomes} (that is, mean and conditional variance calculation), it is standard to verify that for each $t$, 
\begin{align*}
\max_{d \geq 0,x\in\mathcal{X},a\in\mathcal{A}} \Big|\hat{\rho}_t(d|x,a) - \rho_t^\dagger(d|x,a)\Big| = \op(1),
\end{align*}
which further implies that
\begin{align*}
\sup_{e_{t+1},\dots,e_T} \Big|\hat{\mathsf{V}}_t^{\mathtt{forward}}(e_{t+1},\dots,e_{T}|x) - \mathsf{V}_t^\dagger(e_{t+1},\dots,e_T|x)\Big| = \op(1).
\end{align*}
Consistency of the treatment allocation then follows from standard M-estimation argument. See for example \cite{van2000asymptotic}.

\subsection{Proof of Theorem \ref{SA-thm-7: Optimized variance and bound on efficiency loss}}

To show the first claim, fix some $t$ and consider the objective function 
\[
\mathsf{V}^\dagger_t(e_{t+1}(\cdot),\dots,e_T(\cdot)) = \mathsf{V}^\dagger_t\Big(e_{t+1}(\cdot),\dots,e_T(\cdot)\ ;\  \rho^{\dagger}_t(\cdot|\cdot,\cdot)\Big).
\]
Here, we have augmented the notation to emphasize that the objective function depends on the delay mechanism learned at stage $t$. In addition, we remark that the objective function is monotonically decreasing in the delay mechanism. At stage $t$, the optimization process implies that
\[
\min_{e_{t+1}(\cdot),\dots,e_T(\cdot)}\mathsf{V}^\dagger_t\Big(e_{t+1}(\cdot),\dots,e_T(\cdot)\ ;\  \rho^{\dagger}_t(\cdot|\cdot,\cdot)\Big) = \min_{e_{t+2}(\cdot),\dots,e_T(\cdot)}\mathsf{V}^\dagger_t\Big(e_{t+1}^\dagger(\cdot),e_{t+2}(\cdot),\dots,e_T(\cdot)\ ;\  \rho^{\dagger}_t(\cdot|\cdot,\cdot)\Big).
\]
Here, the meaning of the right-hand side is that we first plug in the optimal allocation for stage $t+1$, and then optimize with respect to all future stages. To make progress, we use the fact that conservative extrapolation is employed. Under the conservative extrapolation, the learned delay mechanism never decreases; that is, 
\[
\rho^{\dagger}_t(d|x,a) \leq \rho^{\dagger}_{t+1}(d|x,a)
\]
for all $d$, $x$, and $a$. As a result, we have
\[
\mathsf{V}^\dagger_t\Big(e_{t+1}^\dagger(\cdot),e_{t+2}(\cdot),\dots,e_T(\cdot)\ ;\  \rho^{\dagger}_t(\cdot|\cdot,\cdot)\Big) \geq \mathsf{V}^\dagger_t\Big(e_{t+1}^\dagger(\cdot),e_{t+2}(\cdot),\dots,e_T(\cdot)\ ;\  \rho^{\dagger}_{t+1}(\cdot|\cdot,\cdot)\Big).
\]
The right-hand side coincides with the objective function at stage $t+1$:
\[
\mathsf{V}^\dagger_{t+1}(e_{t+2}(\cdot),\dots,e_T(\cdot)) = \mathsf{V}^\dagger_{t+1}\Big(e_{t+2}(\cdot),\dots,e_T(\cdot)\ ;\  \rho^{\dagger}_{t+1}(\cdot|\cdot,\cdot)\Big) = \mathsf{V}^\dagger_t\Big(e_{t+1}^\dagger(\cdot),e_{t+2}(\cdot),\dots,e_T(\cdot)\ ;\  \rho^{\dagger}_{t+1}(\cdot|\cdot,\cdot)\Big).
\]
This confirms the inequalities
\[
\min_{e_{T}(\cdot)}\mathsf{V}_{T-1}^\dagger(e_{T}(\cdot)) \leq \min_{e_{T-1}(\cdot),e_{T}(\cdot)}\mathsf{V}_{T-2}^\dagger(e_{T-1}(\cdot),e_{T}(\cdot)) \leq \min_{e_{2}(\cdot),\dots,e_{T}(\cdot)}\mathsf{V}_{1}^\dagger(e_{2}(\cdot),\dots,e_{T}(\cdot)).
\]
To conclude the proof of this claim, we notice that $\rho^{\dagger}_{T}(d|\cdot,\cdot) = \rho(d|\cdot,\cdot)$ for all $d\leq T-1$. In other words, by the time the experiment concludes, all relevant information about the delay mechanism has been acquired. As a result, 
\begin{align*}
	\min_{e_{T}(\cdot)}\mathsf{V}_{T-1}^\dagger(e_{T}(\cdot)) &= \mathsf{V}_{T-1}^\dagger(e_{T}^\dagger(\cdot)) =  \mathsf{V}_{T-1}^\dagger\Big(e_{T}^\dagger(\cdot)\ ;\ \rho^{\dagger}_{T-1}(\cdot|\cdot,\cdot)\Big) \\
&\geq \mathsf{V}_{T-1}^\dagger\Big(e_{T}^\dagger(\cdot)\ ;\ \rho^{\dagger}_{T}(\cdot|\cdot,\cdot)\Big) = \mathsf{V}_{T-1}^\dagger\Big(e_{T}^\dagger(\cdot)\ ;\ \rho(\cdot|\cdot,\cdot)\Big).
\end{align*}
In the above, the first equality follows from the definition that $e_{T}^\dagger$ is the optimized allocation; the second equality is simply a notational augmentation to emphasize the dependence on the delay mechanism; the inequality follows from the monotonicity of the objective function with respect to the learned delay mechanism; and the final equality follows from our earlier discussion. It is worth mentioning that $\mathsf{V}_{T-1}^\dagger(e_{T}^\dagger(\cdot)\ ;\ \rho(\cdot|\cdot,\cdot)) = \mathsf{V}$. Clearly, 
\[
\mathsf{V}_{T-1}^\dagger\Big(e_{T}^\dagger(\cdot)\ ;\ \rho(\cdot|\cdot,\cdot)\Big) \geq  \min_{e_{2}(\cdot),\dots,e_{T}(\cdot)}\mathsf{V}_{1}^\dagger\Big(e_{2}(\cdot),\dots,e_{T}(\cdot)\ ;\ \rho(\cdot|\cdot,\cdot)\Big) \geq \mathsf{V}^*.
\]

To show the second claim, we first fix some $\epsilon > 0$ and $d_\epsilon$ such that
\begin{align*}
    \rho(d|x,a) - \rho(d_{\epsilon}|x,a) \leq \epsilon
\end{align*}
for all $x$, $a$, and all $d \geq d_\epsilon$. Then for all $t \geq d_\epsilon$,
\begin{align*}
& \max_{d\geq 0,x\in\mathcal{X},a\in\mathcal{A}}\Big|\rho(d|x,a) - \rho^\dagger_t(d|x,a)\Big| = \max_{d\geq 0,x\in\mathcal{X},a\in\mathcal{A}}\Big(\rho(d|x,a) - \rho^\dagger_t(d|x,a)\Big) \\
\leq& \max_{d\leq t-1,x\in\mathcal{X},a\in\mathcal{A}}\Big(\rho(d|x,a) - \rho^\dagger_t(d|x,a)\Big) + \max_{d\geq t,x\in\mathcal{X},a\in\mathcal{A}}\Big(\rho(d|x,a) - \rho^\dagger_t(d|x,a)\Big)\\
=& 0 + \max_{d\geq t,x\in\mathcal{X},a\in\mathcal{A}}\Big(\rho(d|x,a) - \rho^\dagger_t(d|x,a)\Big)\\
\leq& \max_{d\geq t,x\in\mathcal{X},a\in\mathcal{A}}\Big(\rho(d|x,a) - \rho^\dagger_t(d_\epsilon|x,a)\Big) \leq \epsilon.
\end{align*} 
Fix $t = d_\epsilon$. Then since the objective function is Lipschitz continuous in the learned delay mechanism (as the denominators are bounded away from zero), one has
\begin{align*}
\mathsf{V}^\dagger_t\Big(e_{t+1}(\cdot),\dots,e_T(\cdot)\ ;\  \rho(\cdot|\cdot,\cdot)\Big) \leq \mathsf{V}^\dagger_t\Big(e_{t+1}(\cdot),\dots,e_T(\cdot)\ ;\  \rho^{\dagger}_t(\cdot|\cdot,\cdot)\Big) \leq \mathsf{V}^\dagger_t\Big(e_{t+1}(\cdot),\dots,e_T(\cdot)\ ;\  \rho(\cdot|\cdot,\cdot)\Big) + C\epsilon
\end{align*}
for some constant $C$. To complete the proof, we further augment the notation by considering
\begin{align*}
\mathsf{V}^\dagger_t\Big(e_{t+1}(\cdot),\dots,e_T(\cdot)\ ;\  \rho(\cdot|\cdot,\cdot)\Big) = \mathsf{V}^\dagger_t\Big(e_{t+1}(\cdot),\dots,e_T(\cdot)\ ;\  e_{1}^\dagger(\cdot),\dots,e_{t}^\dagger(\cdot),\rho(\cdot|\cdot,\cdot)\Big),
\end{align*}
to emphasize that the objective function at the end of stage $t$ depends on past allocations, $e_{1}^\dagger(\cdot),\dots,e_{t}^\dagger(\cdot)$. Again, due to the Lipschitz continuity of the objective function, we have
\begin{align*}
\mathsf{V}^\dagger_t\Big(e_{t+1}(\cdot),\dots,e_T(\cdot)\ ;\  e_{1}^\dagger(\cdot),\dots,e_{t}^\dagger(\cdot),\rho(\cdot|\cdot,\cdot)\Big) \leq \mathsf{V}^\dagger_t\Big(e_{t+1}(\cdot),\dots,e_T(\cdot)\ ;\  e_{1}^*(\cdot),\dots,e_{t}^*(\cdot),\rho(\cdot|\cdot,\cdot)\Big)    +C'\frac{d_\epsilon}{T}
\end{align*}
for some possibly different constant $C'$. The second claim of the theorem then follows from the observation that
\begin{align*}
\min_{e_{t+1}(\cdot),\dots,e_T(\cdot)}\mathsf{V}^\dagger_t\Big(e_{t+1}(\cdot),\dots,e_T(\cdot)\ ;\  e_{1}^*(\cdot),\dots,e_{t}^*(\cdot),\rho(\cdot|\cdot,\cdot)\Big) = \mathsf{V}^*.
\end{align*}

\bibliographystyle{jasa}
\bibliography{reference}